\documentclass[structabstract]{aa}  
\usepackage{graphicx}
\usepackage{lscape}
\usepackage{txfonts}
\usepackage{natbib}
\usepackage{sidecap}
\usepackage{comment}
\usepackage{color}
\usepackage{booktabs}
\usepackage[caption=false]{subfig} 

\begin{document}

\title{Massive open star clusters using the VVV survey}
\subtitle{V. Young clusters with an OB stellar population\thanks{Based on observations taken within the ESO VISTA Public Survey VVV 
(programme ID 179.B-2002), and with ISAAC/VLT (programme 087.D-0341(A)).}}
\author{S. Ram\'irez Alegr\'ia\inst{1,2} \and J. Borissova\inst{2,1} \and A.-N. Chen\'e\inst{3} \and C. Bonatto\inst{4} \and R. Kurtev\inst{2,1} 
\and \\ P. Amigo\inst{2} \and M. Kuhn\inst{2} \and M. Gromadzki\inst{1,2} \and J.A. Carballo-Bello\inst{1,2}}

\institute{Millenium Institute of Astrophysics, Av. Vicu\~na Mackenna 4860, 782-0436, Macul, Santiago, Chile \and Instituto de F\'isica y Astronom\'ia, Facultad de Ciencias, Universidad de Valpara\'iso, Av. Gran Breta\~na 1111, Playa Ancha, Casilla 5030, Valpara\'iso, Chile \email{sebastian.ramirez@uv.cl} \and Gemini Observatory, Northern Operations Center, 670 North A'ohoku Place, Hilo, HI 96720, USA \and Universidade Federal do Rio Grande do Sul, Departamento de Astronomia CP 15051, RS, Porto Alegre, 91501-970, Brazil}
\date{Received May 2015 / Accepted February 2016}

\abstract
{The ESO public survey VISTA Variables in the V\'ia L\'actea (VVV) has contributed with deep multi-epoch photometry of the Galactic
bulge and the adjacent part of the disk over 526 square degrees. More than a hundred cluster candidates have been reported thanks to this survey.}
{We present the fifth article in a series of papers focused on young and massive clusters discovered in the VVV
survey. In this paper, we present the physical characterization of five clusters with a spectroscopically confirmed OB-type 
stellar population.}
{To characterize the clusters, we used near-infrared photometry ($J$, $H,$ and $K_S$) from the VVV survey 
and near-infrared $K$-band spectroscopy from ISAAC at VLT, following the methodology presented in the previous articles 
of the series.}
{All clusters in our sample are very young (ages between 1-20 Myr), and their total mass are between $(1.07^{+0.40}_{-0.30})\cdot10^2$
 $M_{\odot}$ and $(4.17^{+4.15}_{-2.08})\cdot10^3$ $M_{\odot}$. We observed a relation between the clusters total mass $M_{ecl}$
 and the mass of their most massive stellar member $m_{max}$, even for clusters with an age $<$ 10 Myr.}
{}

 \titlerunning{VVV open star clusters V}
\maketitle


\section{Introduction}\label{intro}

Young massive clusters (cluster total mass $M>10^4  M_{\odot}$, \citealt{portegies10} or $M>10^3  M_{\odot}$, \citealt{hanson07}) 
play a central role in the structure and evolution of their host galaxies. The clusters' massive stellar population may induce or diminish the 
formation of other stars, add kinetic energy to the interstellar medium through their massive winds, and enrich the medium with metallic 
elements during their lives through stellar winds and in their final explosion as supernova. The young clusters also help us to trace the 
spiral structure of the Milky Way and to study massive stellar population in several evolutionary stages. 

Despite their massive and luminous population, massive clusters are located behind a large amount of extinction. Clusters younger 
than a few Myr, classified as Phase I according to \citet{portegies10}, present on-going stellar formation and a large amount of gas and dust.
The dust often makes their detection impossible when using optical images. This is one of the main reasons that we know of less than 
10~\% of the expected population. Near-infrared all-sky surveys -- such as the Two Micron All-Sky Survey, \citep[2MASS,][]{skrutskie06}, 
the Galactic Legacy Infrared Midplane Survey Extraordinaire, \citep[GLIMPSE,][]{benjamin03}, and the UKIRT Infrared Deep 
Sky Survey \citep[UKIDSS,][]{lawrence07}-- made it possible to discover the most massive stellar clusters in the Galaxy. 

 The ESO public survey VISTA Variables in the V\'ia L\'actea \citep[VVV,][]{minnitiVVV10,saito10,saito12} is a perfect tool for this exploration. 
 The VVV survey covers the Galactic bulge and the adjacent disk region, including the far edge of the bar, with a spatial 
 resolution of 0.34 arcsec pix$^{-1}$ in \textit{ZYJH$K_S$} filters. Three catalogues with new cluster candidates have been published
 using VVV data: \citet{borissova11}, \citet{borissova14}, and \citet{solin14}. 
 
In this paper we present the characterization of five young clusters from the \citet{borissova11} catalogue. The paper is part of a 
series dedicated to young clusters discovered using the VVV survey \citep{chene12,chene13,chene15,ramirezalegria14}. The methodology 
 for the observation and characterization of the clusters is similar to the one presented in the previous articles of the series. The observations
 are described in Section \ref{method}, together with the general method used for the cluster analysis. In Section \ref{resultados}, we 
 present the individual parameters derived for each cluster. In this section we also discuss the relation proposed by \citet{weidner10} between the cluster 
 total mass and the mass of its most massive star ($M_{ecl} - m_{max}$), using our sample of new characterized clusters. Finally, we 
 summarize the results of this paper in 
 Section \ref{conclusiones}.

\begin{figure*}[t]
\centering
\includegraphics[width=5.7cm,angle=0]{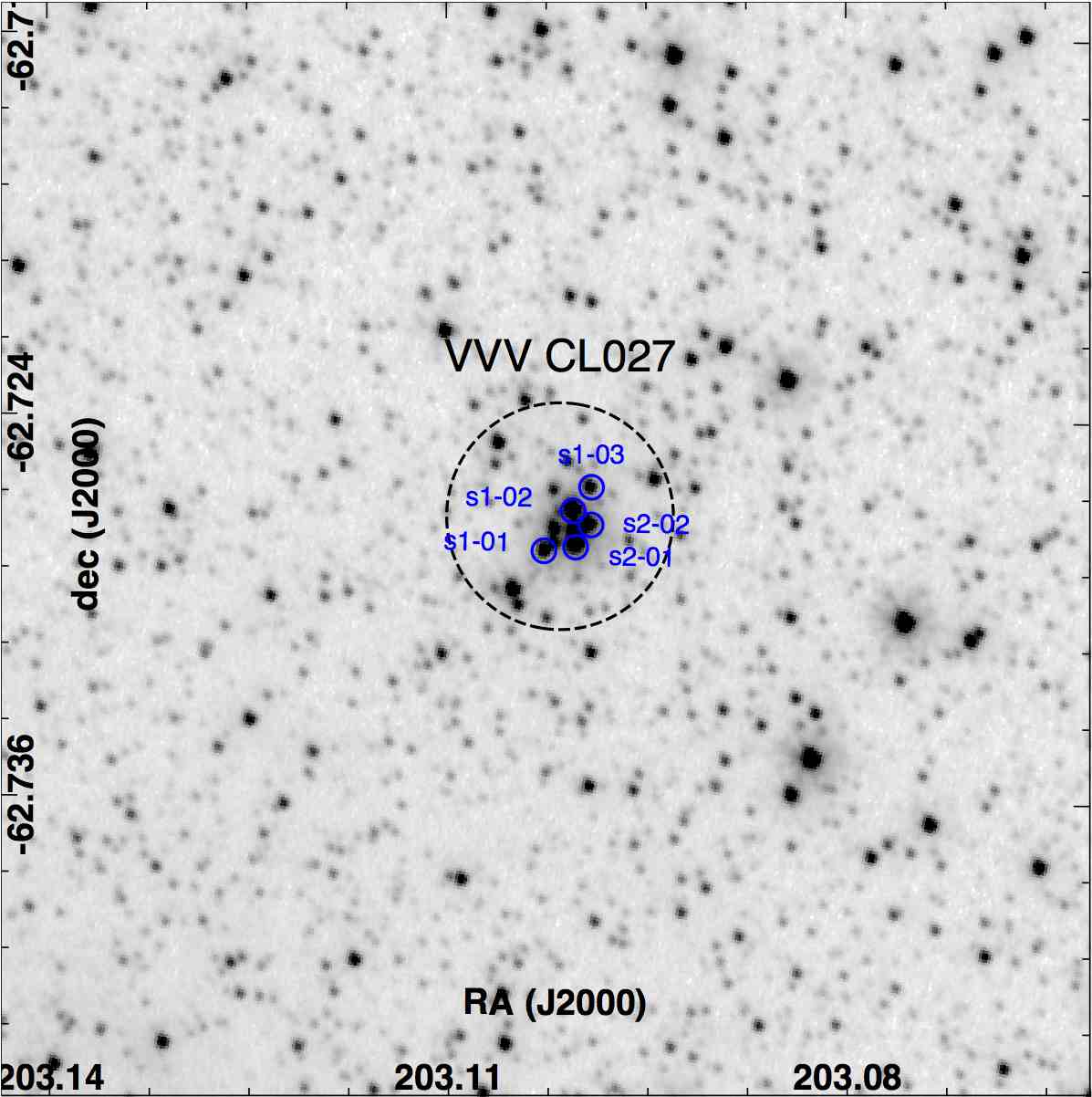}
\vspace{0.1cm}
\includegraphics[width=5.7cm,angle=0]{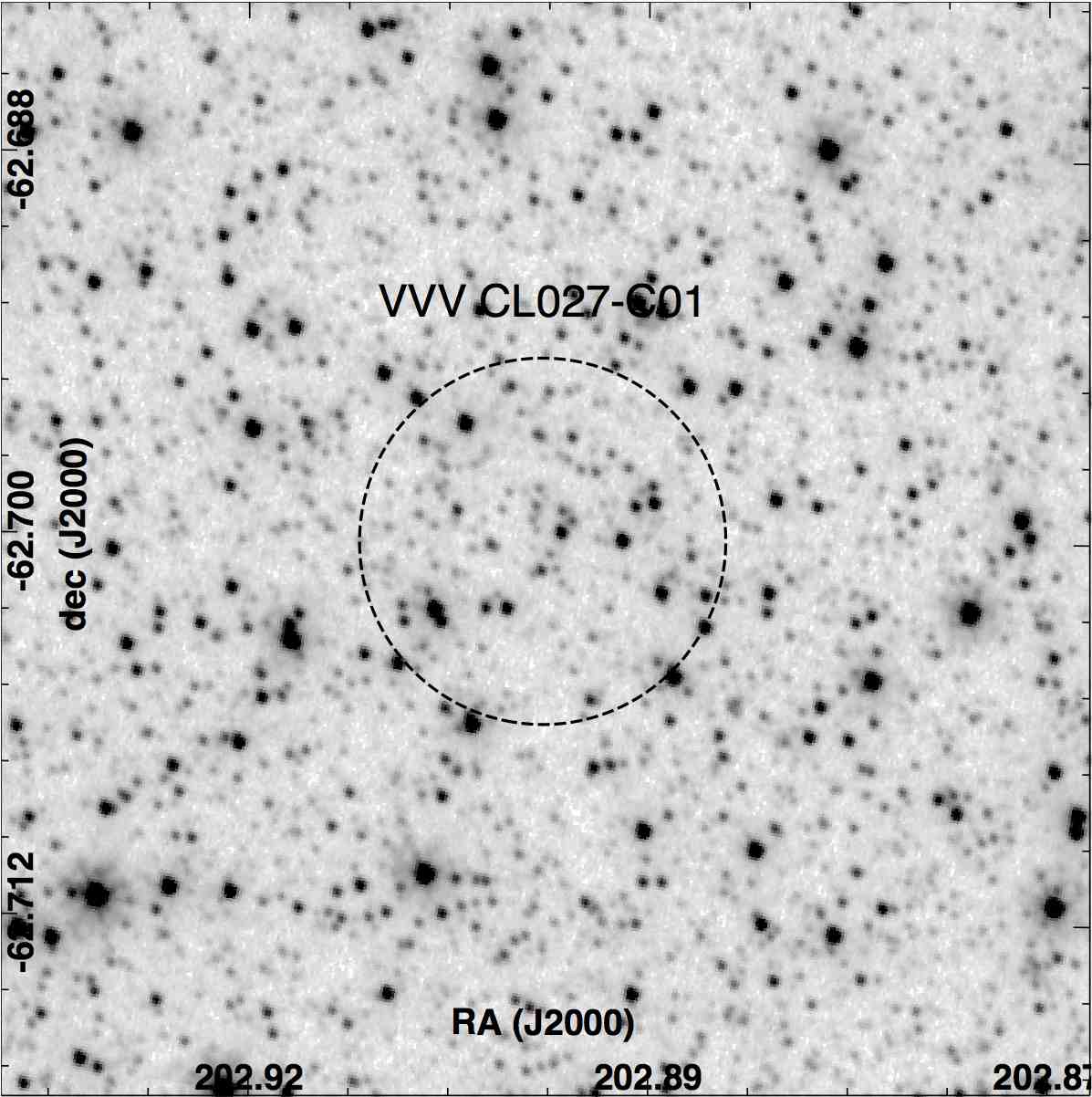}
\includegraphics[width=5.7cm,angle=0]{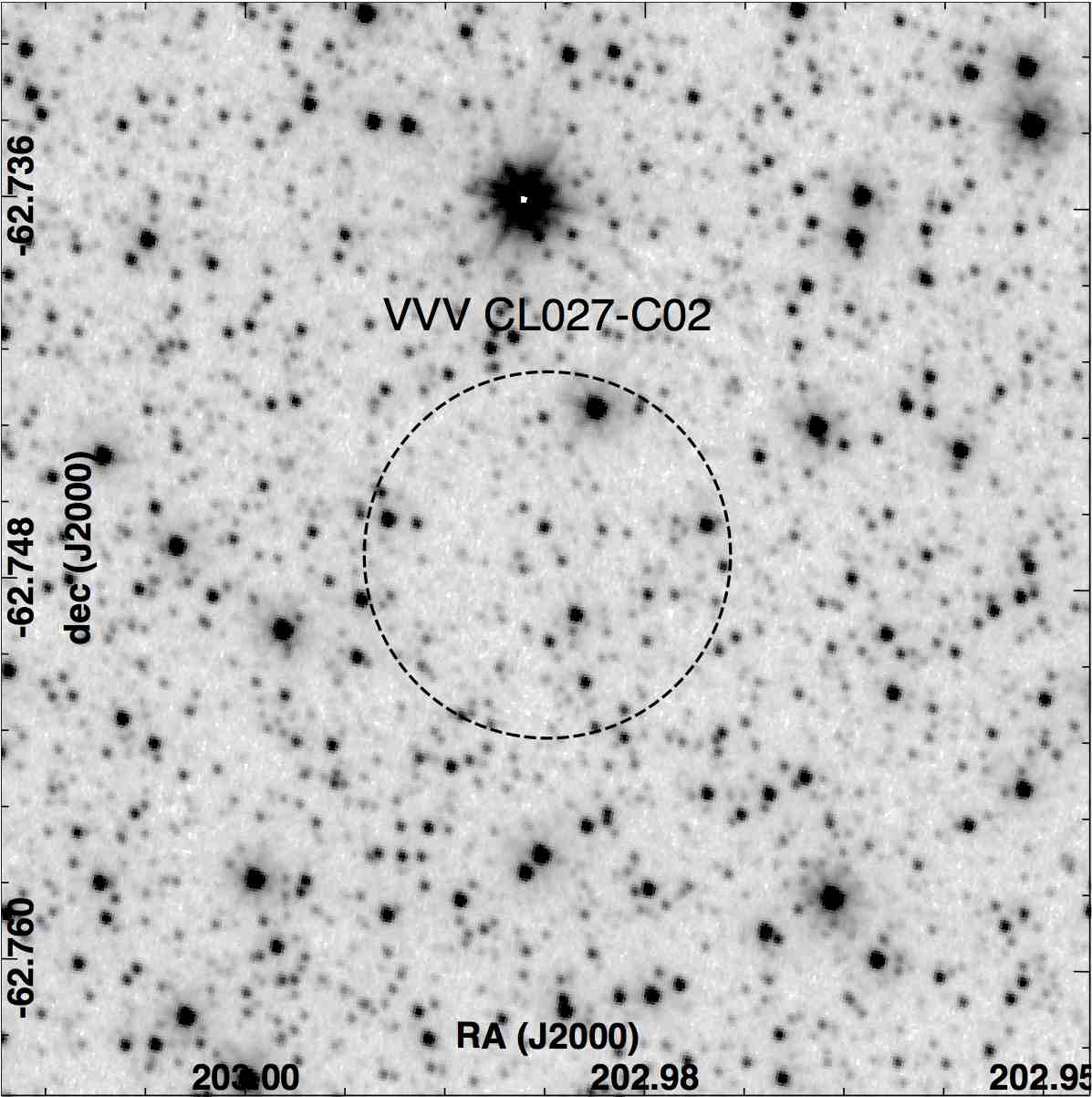}
\vspace{0.1cm}
\includegraphics[width=5.7cm,angle=0]{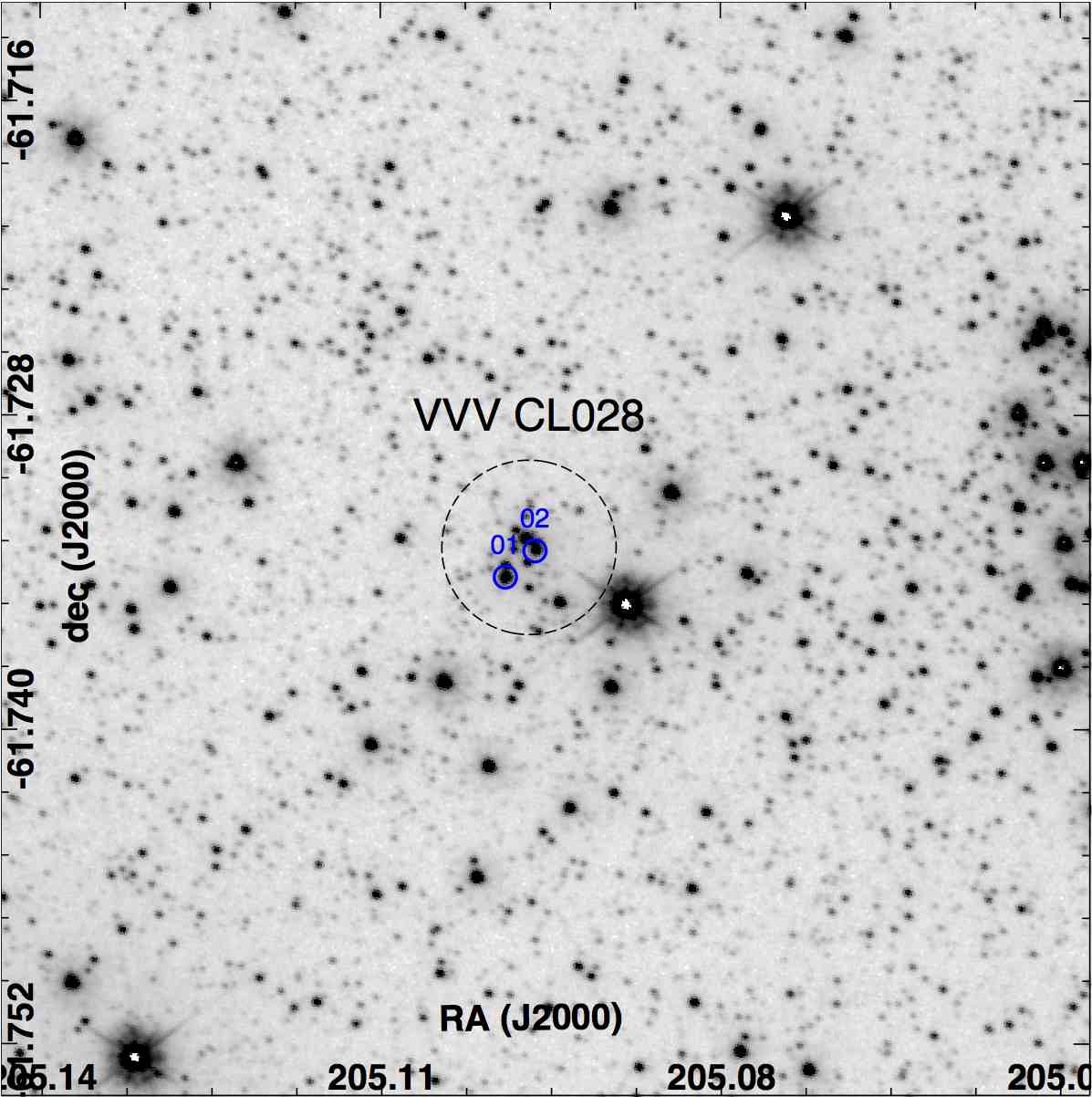}
\includegraphics[width=5.7cm,angle=0]{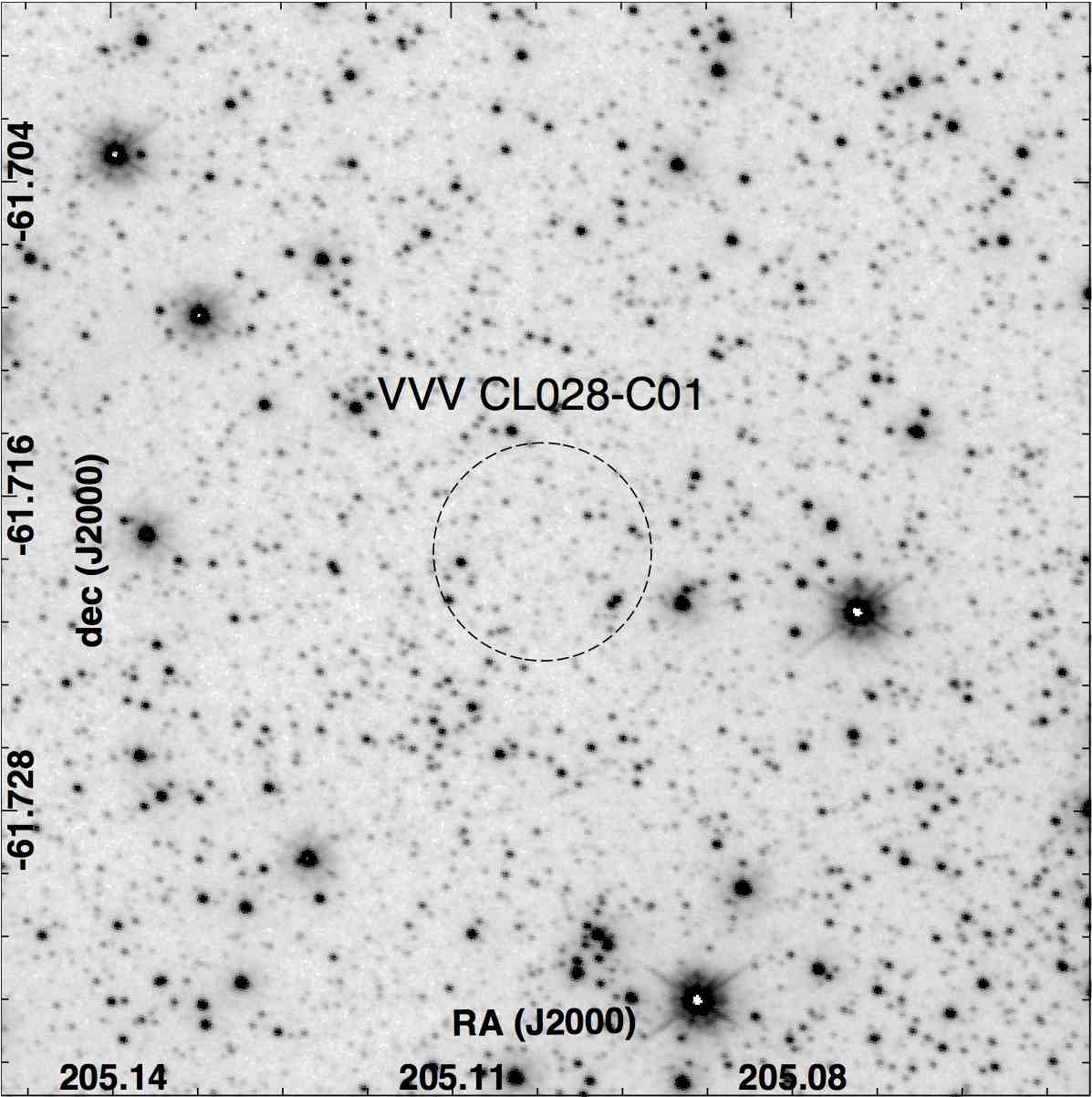}
\includegraphics[width=5.7cm,angle=0]{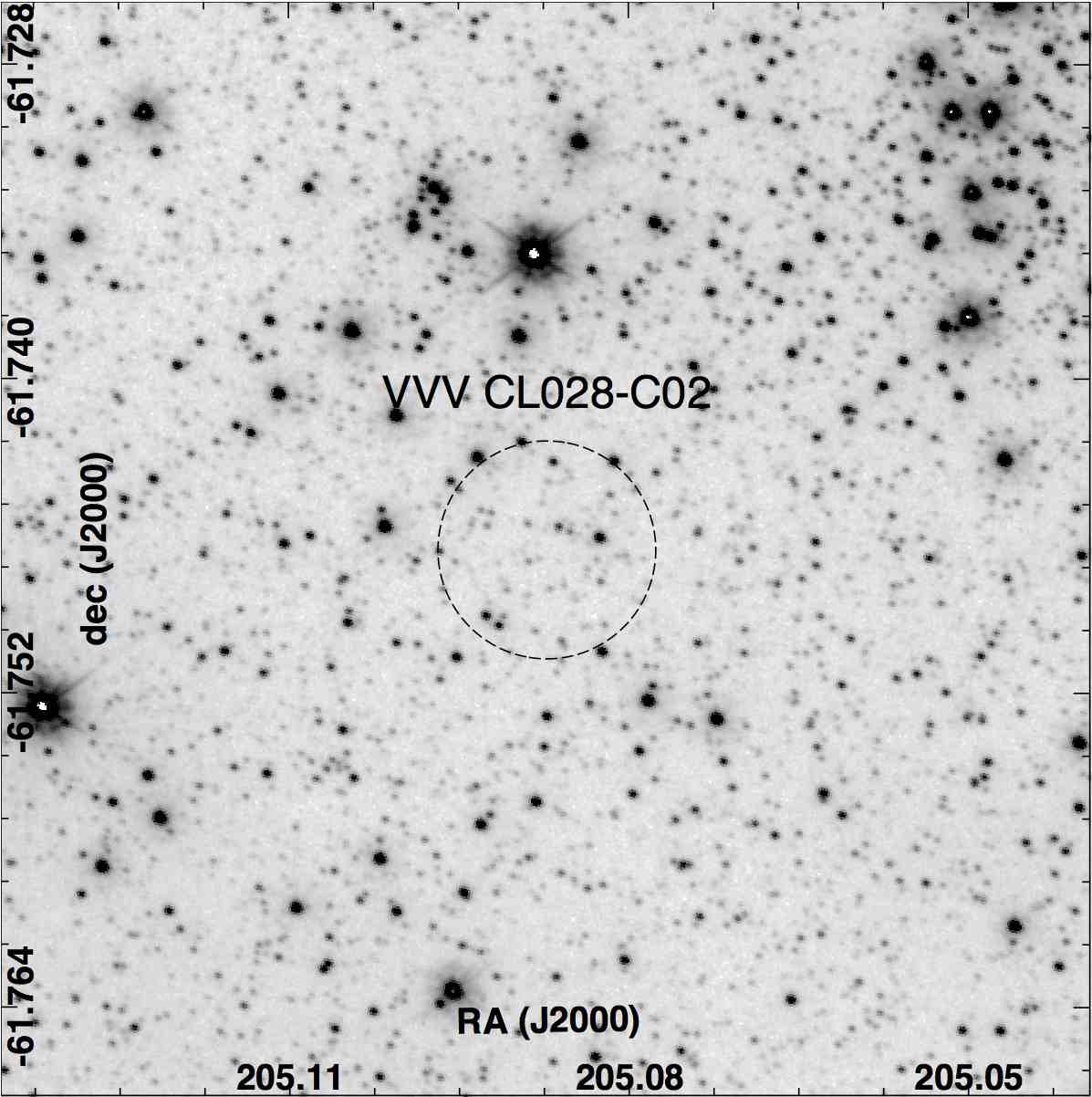}
\vspace{0.1cm}
\includegraphics[width=5.7cm,angle=0]{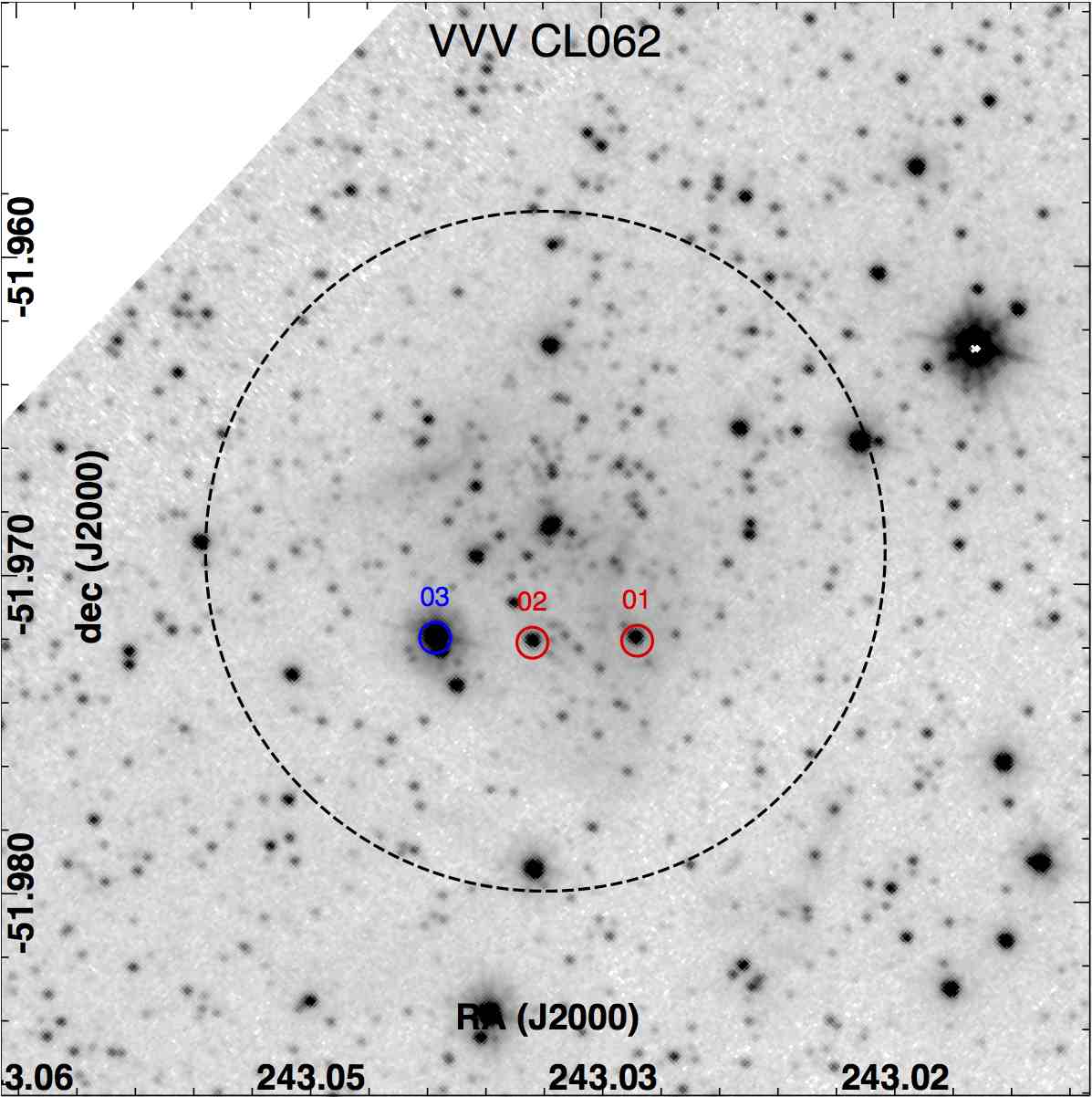}
\includegraphics[width=5.7cm,angle=0]{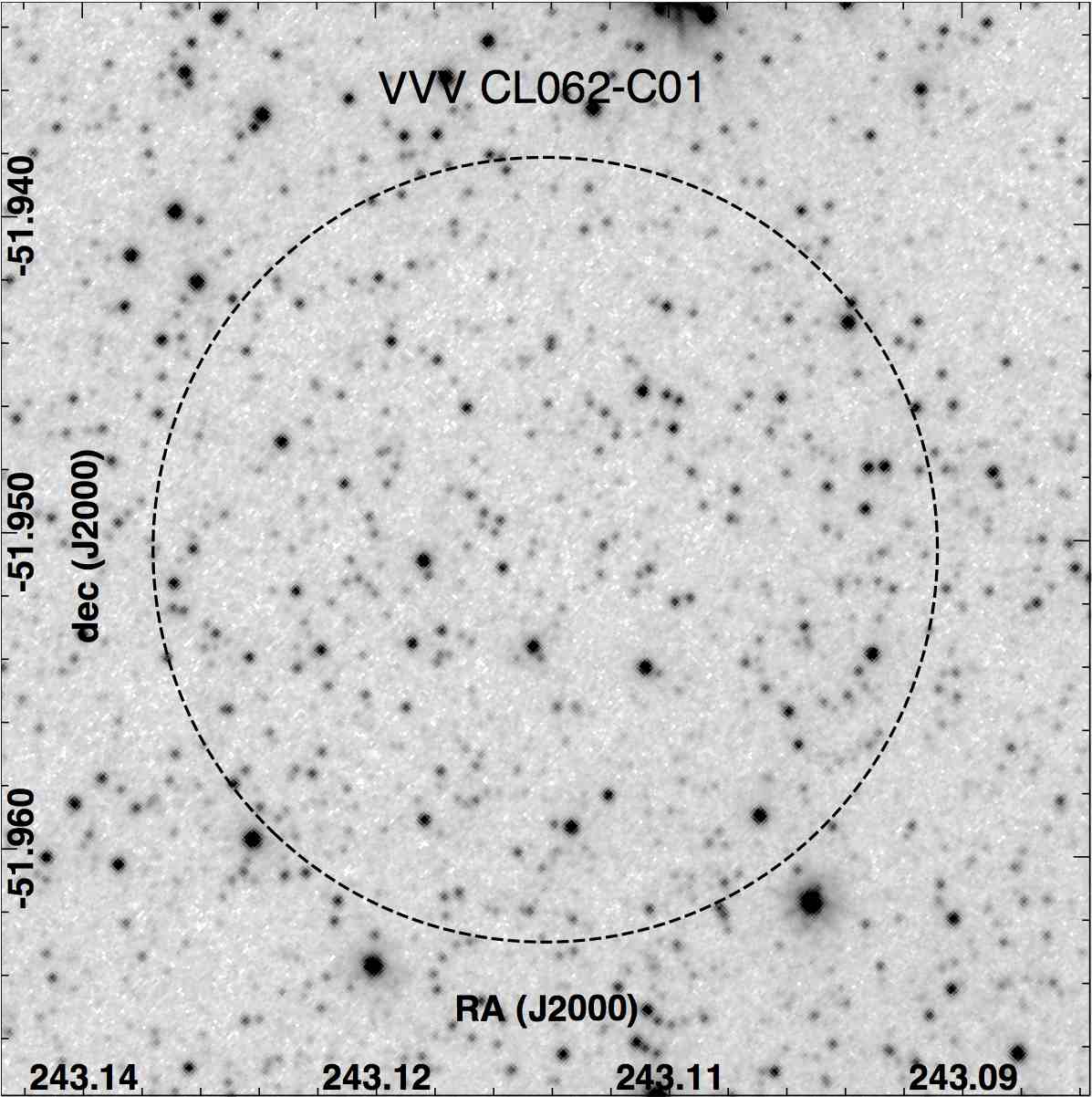}
\includegraphics[width=5.7cm,angle=0]{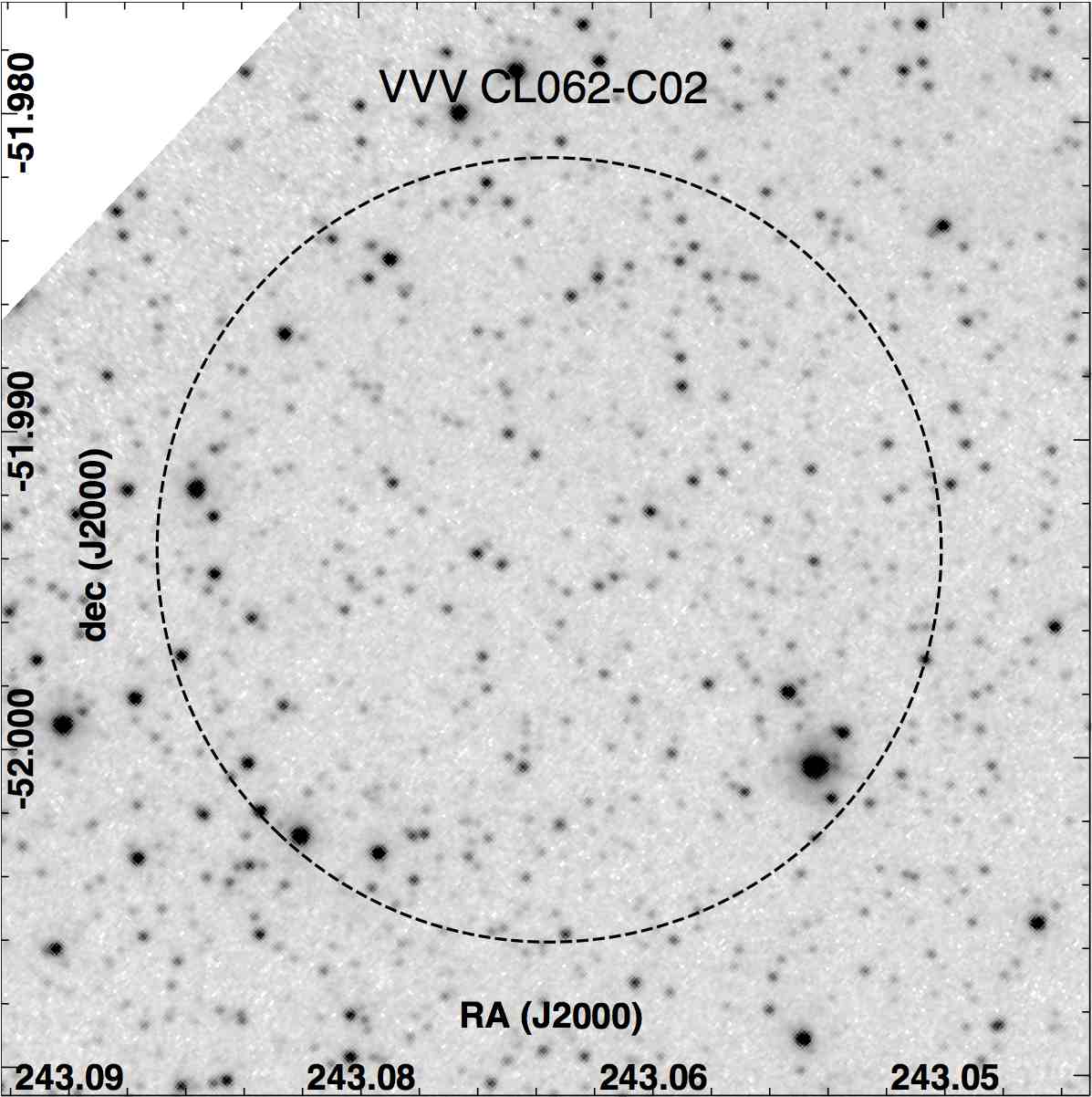}
\includegraphics[width=5.7cm,angle=0]{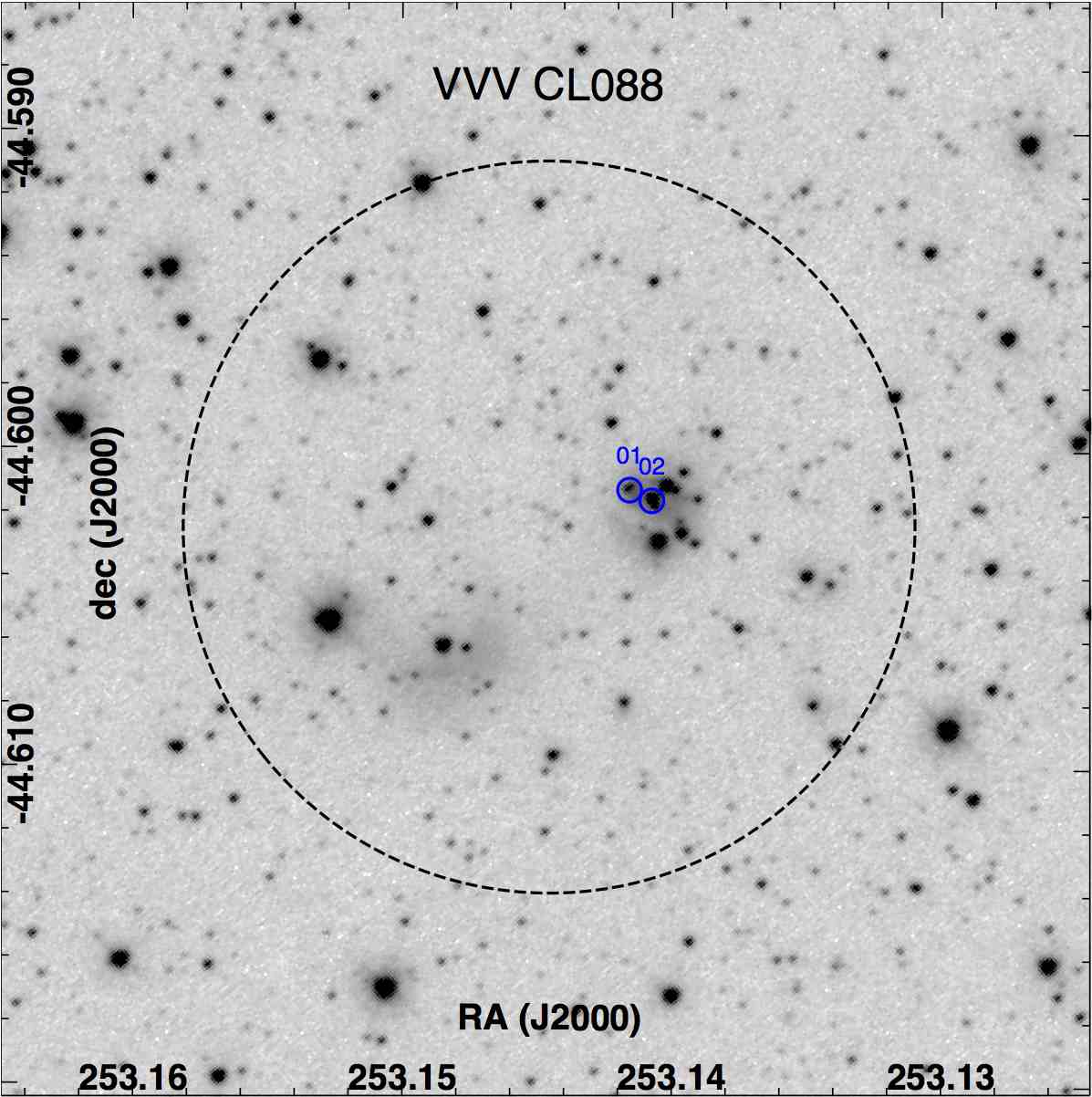}
\includegraphics[width=5.7cm,angle=0]{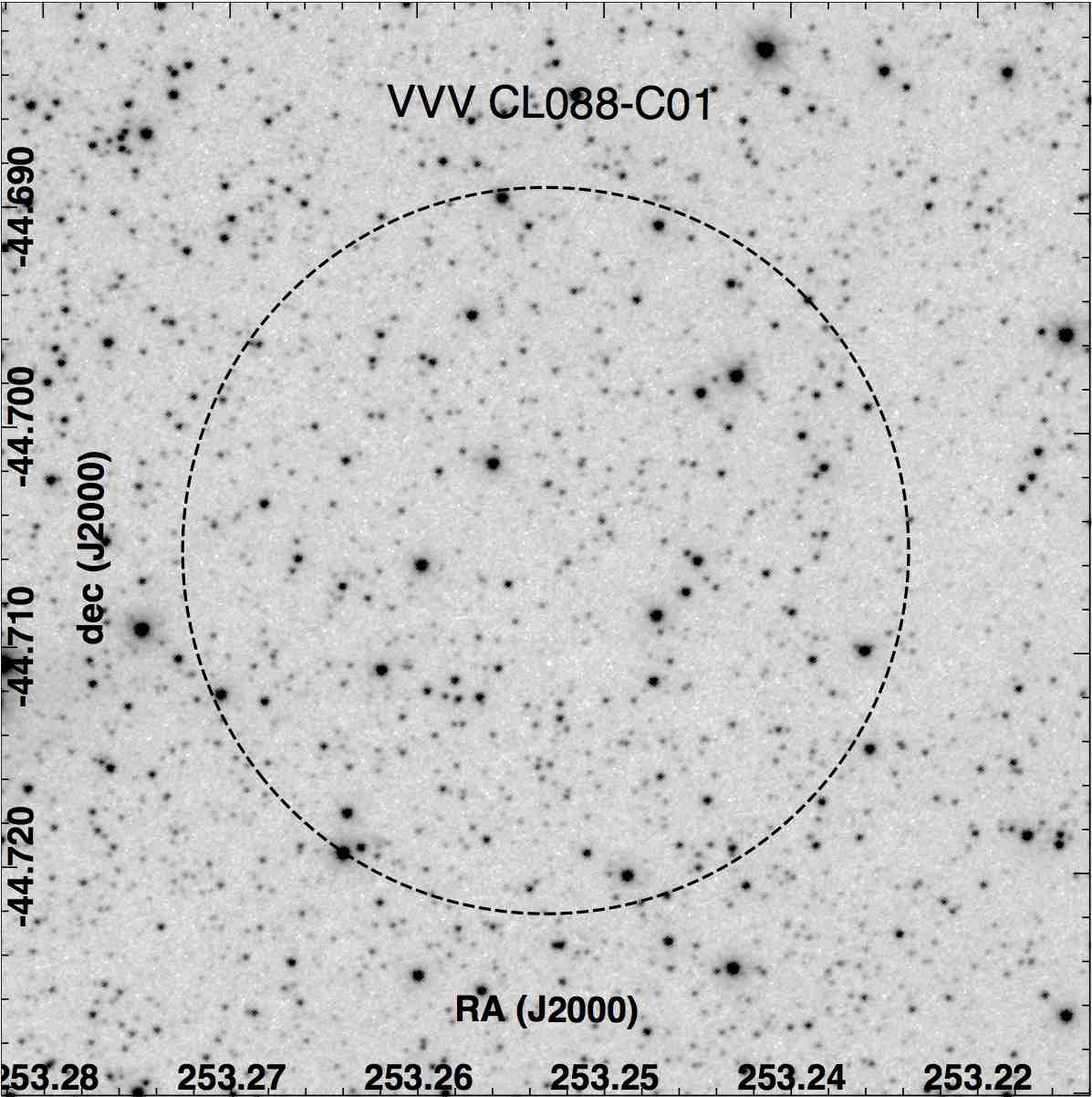}
\includegraphics[width=5.7cm,angle=0]{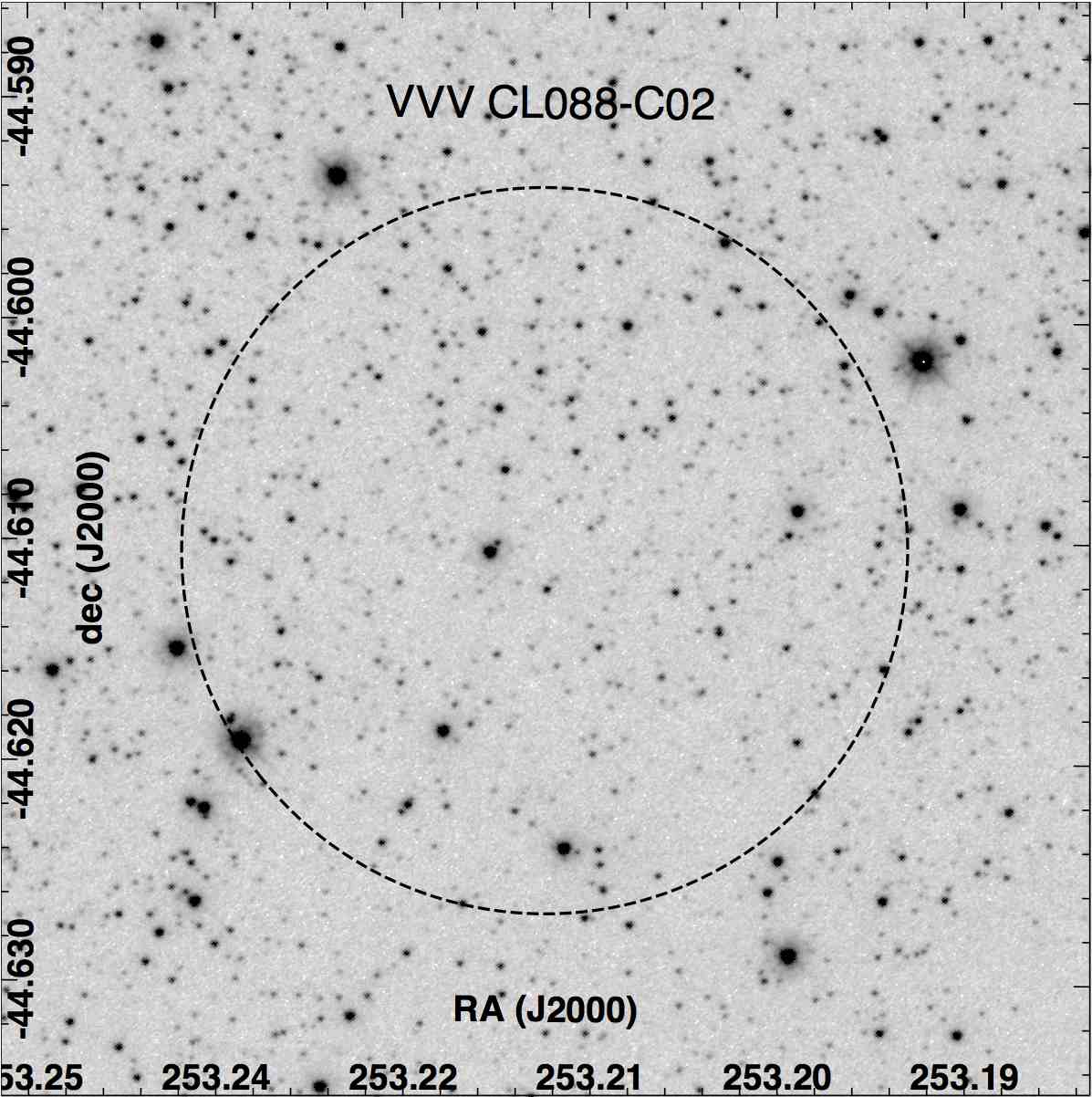}
\end{figure*}

\begin{figure*}
\centering
\includegraphics[width=5.7cm,angle=0]{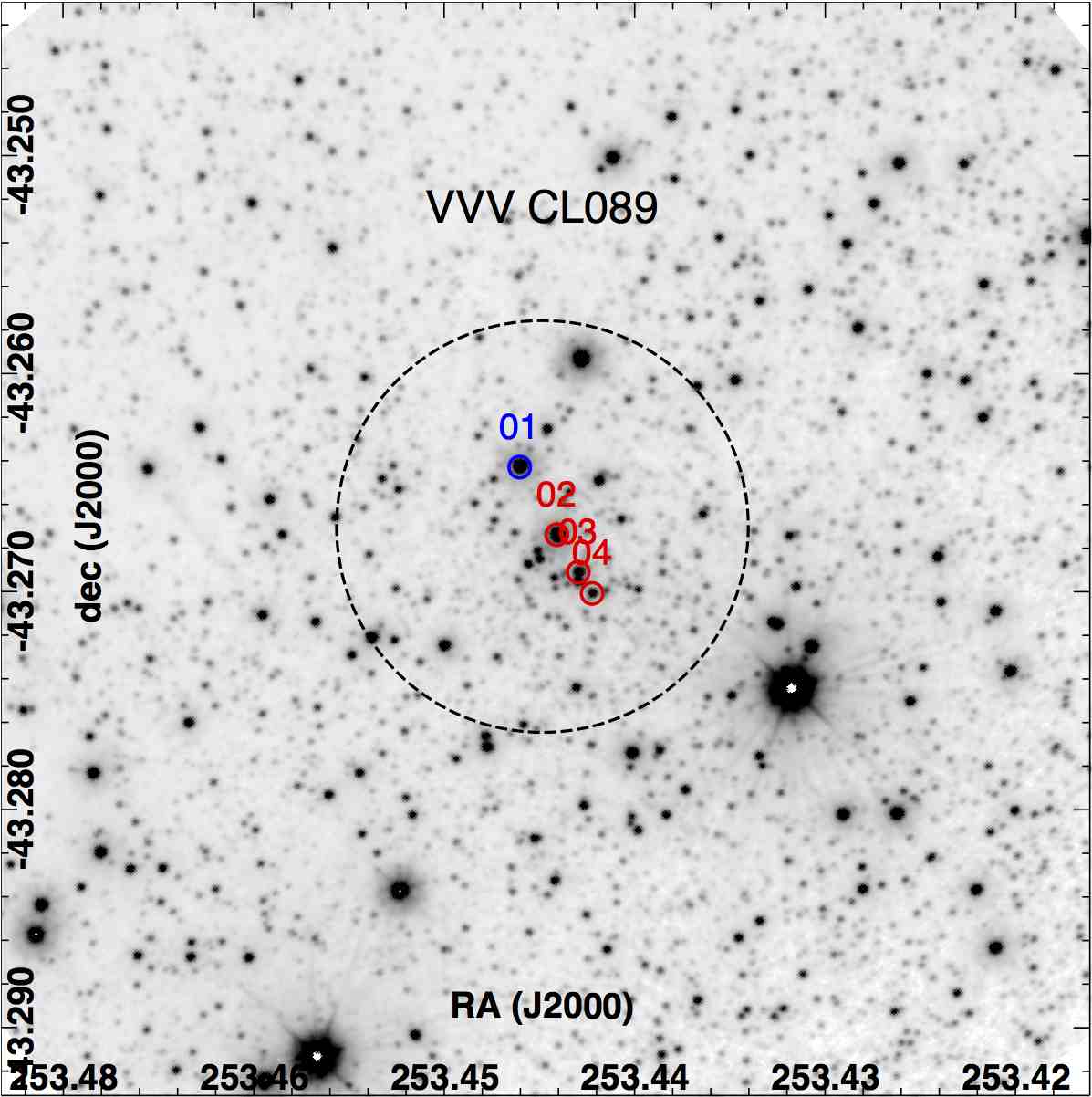}
\includegraphics[width=5.7cm,angle=0]{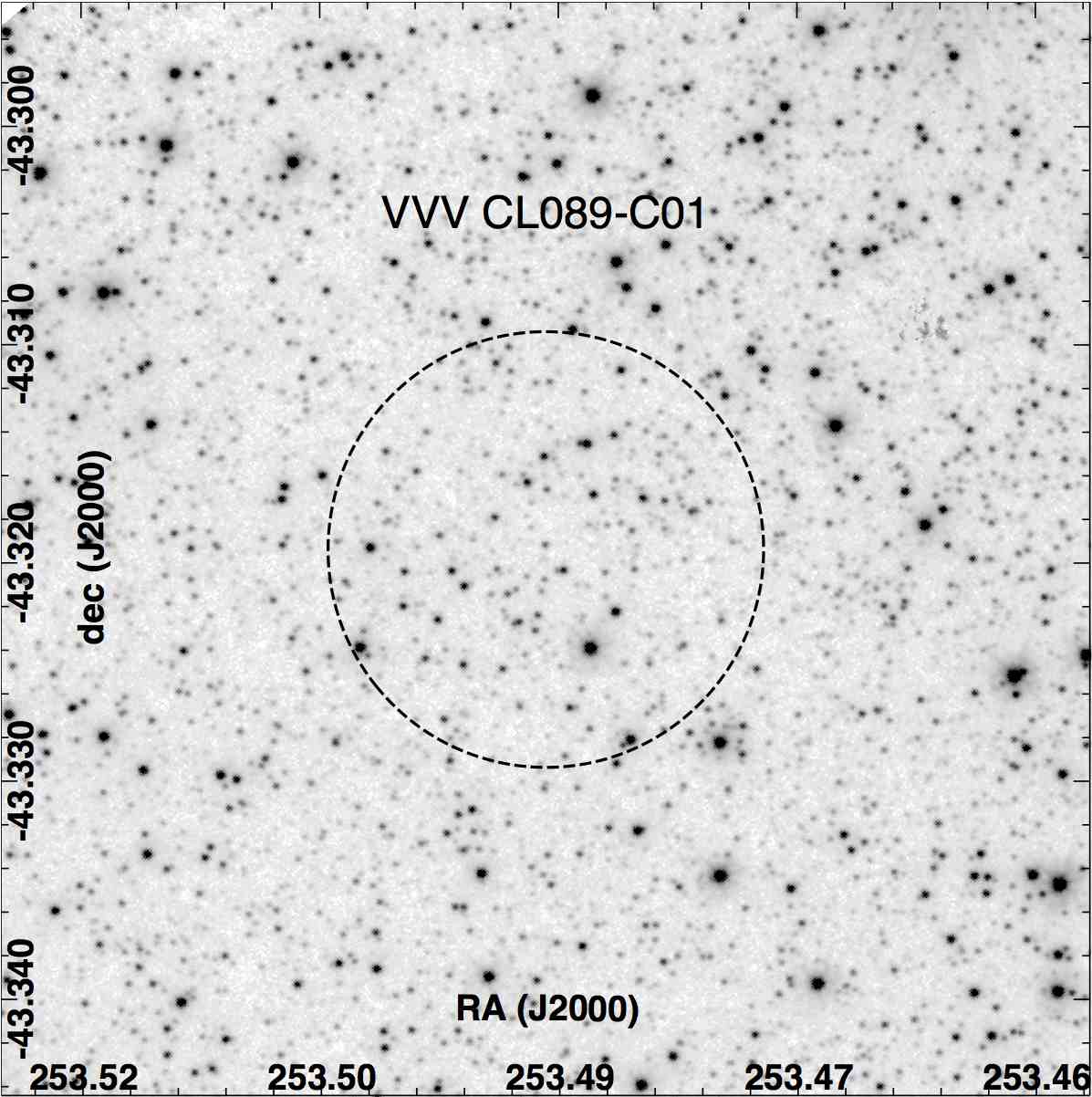}
\includegraphics[width=5.7cm,angle=0]{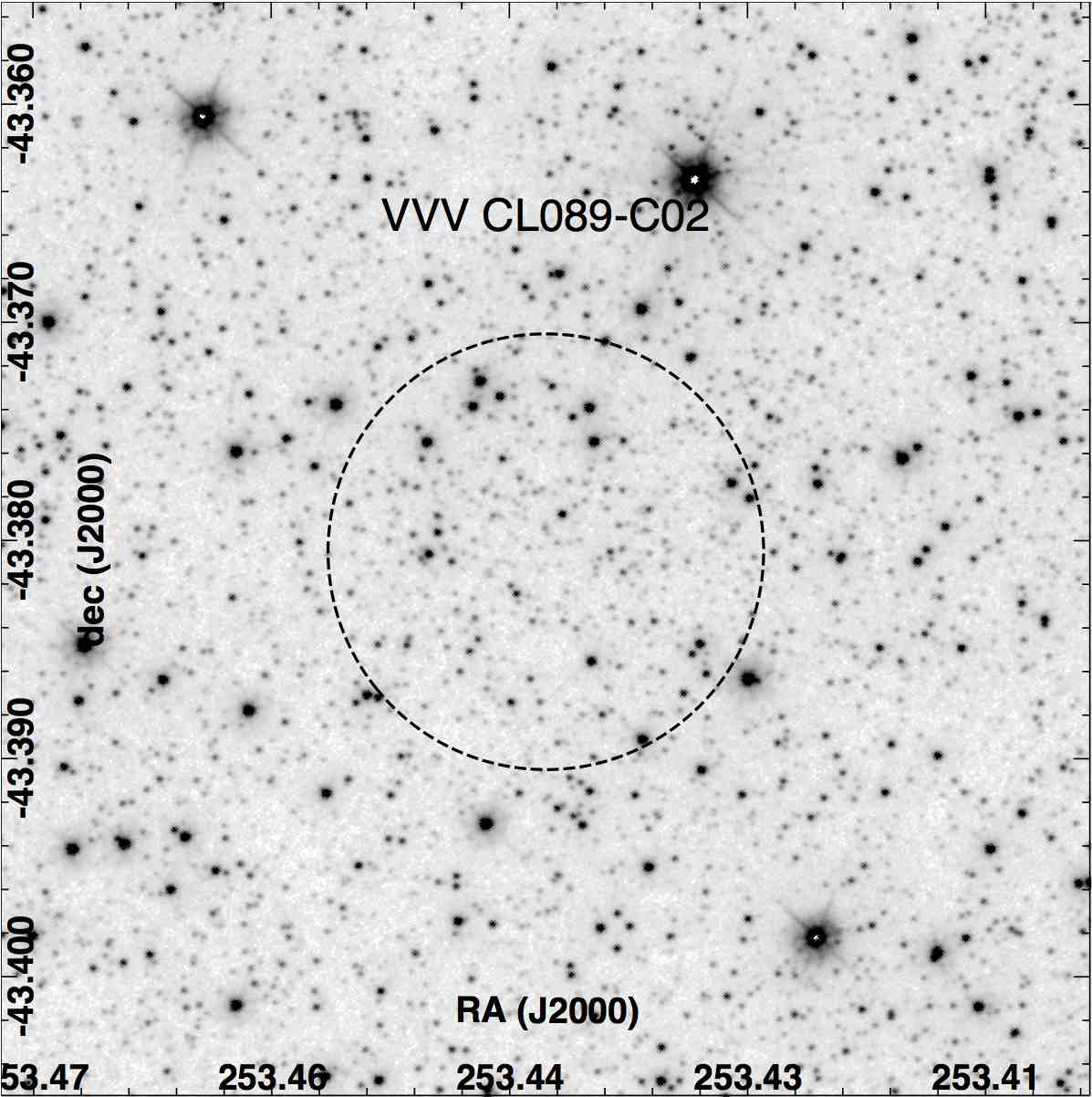}
       \caption{VVV $K_S$ images for the clusters included in this work. The black dashed circles indicate the position and their estimated 
       extensions, while the smaller circles show the positions of the early (blue) and late-type (red) stars. The images 
       are 2 arcmin $\times$ 2 arcmin size. North is up, east to the left.}
       \label{ks_figure}
\end{figure*}
 

\section{Observations and general method}\label{method}

\subsection{Observations}\label{obs}

\begin{table*}
\caption{Clusters with ISAAC spectroscopic follow-up. Equatorial and galactic coordinates are included.}
\begin{center}
\begin{tabular}{cccccc}
\toprule
 Cluster ID & RA (J2000) & Dec (J2000)   &  l        &  b      &   Obs. dates \\
      & [deg]           & [deg]              & [deg]  & [deg]                            &              \\
 \midrule
VVV\,CL027      & 203.100        &  -62.727  &  307.605  & -0.228  & 11 April 2011; 4, 15 May 2011 \\
VVV\,CL028      & 205.096        &  -61.733  &  308.693  & +0.590 & 5 May 2011\\
VVV\,CL062      & 243.033        &  -51.969  &  331.172  & -0.458  & 6 May 2011 \\
VVV\,CL088      & 253.142        &  -44.602  &  341.129  & -0.347  & 23 April 2011 \\
VVV\,CL089      & 253.446        &  -43.267  &  342.301  & +0.329 & 21 April 2011 \\
\bottomrule
\end{tabular}
\end{center}
\label{data_obs}
\end{table*}

For this work we used near infrared images from the VVV survey and spectroscopic data from the Infrared Spectrometer and Array Camera 
(ISAAC) at the Very Large Telescope (VLT, Cerro Paranal, Chile).The studied clusters are part of a larger spectroscopic campaign and were 
selected by the presence of OB-type stars in their population. The original group also included VVV\,CL086 \citep{ramirezalegria14} and 
VVV\,CL041 (Chen\'e et al. 2015, submitted).

The VVV survey uses the VISTA Infrared Camera (VIRCAM) at the Visible and Infrared Survey Telescope for Astronomy (VISTA) 
\citep{emerson06,dalton06}. VIRCAM has a 16-detector array (each detector with a 2048$\times$2048 pixel size), and a pixel scale 
of 0.34 arcsec pix${}^{-1}$. The 16 detectors are separated by gaps (42.5\% of the detector in the Y axis, 90\% in the X axis), forming 
an array called pawprint with an individual coverage of 0.59 square degrees. After a series of six vertical and horizontal shifts, the 
individual pawprints are combined into a single tile with a total of 1.64 square degrees field of view \citep{saito12}.

We measured the photometry directly from the stacked images   from the VISTA Science Archive (VSA) 
website\footnote{http://horus.roe.ac.uk/vsa/}. We used all available images from completed programmes between February 
2010 and August 2013. We used between 32 and 70 images for the clusters in the $K_S$ filter, since VVV survey has 
multi-epoch data in this filter.

We used the VVV-SkZ pipeline \citep{mauro13} to determine the stellar photometry. This is an automated software based on 
ALLFRAME \citep{stetson94} and optimized for VISTA PSF photometry. The pipeline uses only the paw-print images
to avoid the variable PSF observed across the combined tiles. The VVV-SkZ pipeline allows the PSF to vary quadratically across 
the paw-print field of view.

 The final catalogue generated by the pipeline was calibrated astrometrically and photometrically using 2MASS. Photometric errors are 
 lower than 0.2 mag for $K_S < 18.5$ mag and 0.5 mag for $K_S=18.0-20.0$ mag. We assumed 2MASS photometry for 
 sources brighter than $K_S = 11$ mag to avoid saturation.

For near-infrared spectroscopy, we selected bright stars from the cluster region. These stars
were observed with ISAAC/VLT in service mode during April and May 2011. Spectra were 
acquired in the $K$ band ($1.98-2.43 \mu m$), using a 0.3\arcsec wide slit with a resolution $R\sim3000$.
To correct the atmospheric OH emission lines, we nodded along the slit 
following an ABBA nod pattern. The average signal-to-noise ratio per pixel ranges from 
30 to 150. To correct the telluric lines, we observed bright B8 to A2\,V stars (telluric standards). The spectra of these 
objects only presents the Brackett series, which can be modelled and subtracted, leaving only the telluric lines. 
To ensure similar conditions between the cluster stars and the standard telluric lines, we observed them using the 
same airmass and within a reduced time frame.

 For spectroscopic data reduction (flat-fielding, sky subtraction, spectra extraction, and wavelength calibration), we 
used the Interactive Data Language (IDL) and {\sc iraf}\footnote{{\sc iraf} is distributed by the National Optical Astronomy 
Observatories, which are operated by the Association of Universities for Research in Astronomy, Inc., under cooperative 
agreement with the National Science Foundation.} scripts in a similar procedure to the one described by \citet{chene12} and
\citet{chene13}. For the wavelength calibration, we used the OH line from the spectra. This was done because of the sudden 
shifts in the direction of the spectral dispersion along the calibration lamp. The shifts occurred during the observing night, 
making it impossible to determine a unique wavelength solution for the entire run. The final wavelength solution has 
a RMS uncertainty of $\sim$0.5 pixels.

 Figure \ref{ks_figure} shows all the stars that were spectroscopically observed plotted on 2 arcmin $\times$ 2 arcmin 
$K_S$ VVV-image sections. Table \ref{data_obs} summarizes the clusters coordinates (both equatorial and galactic coordinates)
 and the observing dates.

 \subsection{Statistical decontamination}

 The $JHK_S$ photometry was statistically field-star-decontaminated using the FORTRAN algorithm of \citet{bonatto10}, as described by \citet{borissova11}
 and the papers of this series \citep{chene12,chene13,ramirezalegria14}. We used the latest version of the algorithm, which includes polygonal control regions, 
 artificial reddening of the control-field population, and an algorithm for determing grid size in colour-magnitude space. The algorithm divides the $K_S$, $(H-K_S)$ 
 and $(J-K_S)$ ranges  into a grid of cells. In each cell, it estimates the expected number density of cluster stars by subtracting the respective field-star number 
 density. By summing over all the cells, it obtains a total number of member stars, $N_{mem}$. Grid shifts of $\pm$1/3 the cell size are applied in each axis, producing 
 729 independent binnings and $N_{mem}$. The average of these 729 $N_{mem}$ (or $\langle N_{mem}\rangle$) is the limit for considering a star as a 
possible cluster member. Only the $\langle N_{mem}\rangle$ with the highest survival frequency after all tests were finally considered as cluster members. 

To ensure photometric quality, we restricted the analysis to stars with magnitude and colour errors smaller than 0.2 mag. The coordinates of 
the control fields and the radius used for the decontaminations are given in Table \ref{decont_parameters}. The area of the control fields was
corrected to match the cluster field area, and the position and size of the control fields were chosen based on the distribution of stars and clouds 
in the image.

\begin{table}
\caption{Control field parameters for statistical field-star decontamination.}
\begin{center}
\begin{tabular}{cccccc}
\toprule
 ID & RA (J2000) & Dec (J2000)   & Radius \\
      & [deg]          & [deg]               & [arcmin] \\
 \midrule
VVV\,CL027      & 202.899994 &  -62.700001 &  0.35 \\
                         & 202.981827 & -62.746994  & 0.35  \\
VVV\,CL028      & 205.100601 &  -61.718102 &  0.25 \\
                          & 205.087265 &  -61.746513 & 0.25 \\
VVV\,CL062      & 243.111435  & -51.950371  & 0.75  \\
                         & 243.065536 &  -51.993530 &  0.75 \\
VVV\,CL088      & 253.251495 &  -44.705376 &  1.00 \\
                         & 253.214630 &  -44.610310 &  1.00 \\
VVV\,CL089      & 253.485992 &  -43.319336 &  0.60 \\
                         & 253.437851 &  -43.380417 &   0.60  \\

\bottomrule
\end{tabular}
\end{center}
\label{decont_parameters}
\end{table}

 \subsection{Spectral classification}
 
\begin{figure*} 
\centering
\includegraphics[width=6in]{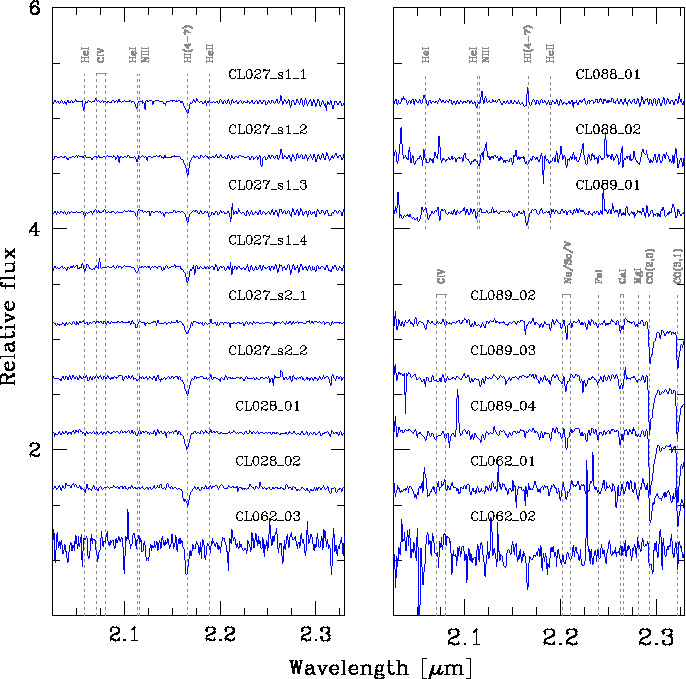} 
            \caption{Individual $K$-band spectra. Spectral features used for the spectral classifications are labelled in grey. 
            Early type spectra are arranged by cluster name, and late type spectra are shown in the bottom of the right column.}
           \label{all_spectra}
\end{figure*}

We based the near-IR spectral classification of the OB stellar types on the two Hanson et al. atlases.
To assure a similar resolution between our spectra and the comparison ones, we 
used only spectra with R$\sim$3000 from the \citet{hanson96} catalogue and degraded the resolution for the \citet{hanson05} 
spectra, using the {\sc iraf} task {\sc splot}. For later spectral types, we used \citet{wallacehinkle97}.

The classification scheme is based on the detection of absorption lines and the comparison of their depth and shape 
with similar resolution spectra of known spectral types. The spectral lines used for the qualitative spectral classification 
are labelled in the individual spectra of Figure \ref{all_spectra}. For early type stars we used the \ion{He}{I} at 2.06 and 
2.11 $\mu m$, \ion{He}{II} at 2.19 $\mu m$, the \ion{H}{I} (4-7) at 2.17 $\mu m$ lines, the \ion{N}{III} multiplet at 2.12 $\mu m$, and the 
\ion{C}{IV} triplet at 2.08 $\mu m$. For the assigned spectral type we assumed an error of $\pm 2$ subtypes, similar to 
\citet{hanson10} and \citet{negueruela10}. Table \ref{data_stars} contains the coordinates, near-infrared magnitudes, 
and spectral types for the stars with spectroscopic observations. 

 \subsection{Cluster physical characterization}
 
  For individual distance estimates, we compared the apparent and intrinsic magnitude derived from the individual spectral 
type. For clusters with two or more stars where the distance estimates agree with a common distance, we assumed 
the averaged individual estimates as the cluster parameter (extinction and distance). We adopted the \citet{stead09} 
extinction law and intrinsic magnitudes from \citet{martins05} in the case of O-type stars. 
We used \citet{cox00} for stars later than B0\,V. Distance errors are dominated by the spectral type determination uncertainty. We estimated them
by deriving the individual distance for the same star assuming $\pm2$ spectral subtypes. The cluster distance and extinction 
errors were estimated using the error propagation method described by \citet{barlow04}. Table \ref{data_stars} includes the 
individual extinction and distance determinations for those stars with spectroscopy.

To estimate the total cluster mass, we first constructed the cluster present-day mass function (P-DMF) using the colour-magnitude diagram
(CMD) and then integrated the Kroupa \citep{kroupa01} initial mass function (IMF) fitted to the cluster IMF. Because we did not 
detect evolved stars in our spectroscopic sample and most of the clusters are surrounded by a nebulosity detected in mid-infrared 
images, we assumed that the P-DMF and the IMF are equivalent.

We obtained the cluster P-DMF by projecting all CMD stars following the reddening vector to the main sequence 
located at the cluster distance. The main sequence is defined by the colours and magnitudes given by \citet{cox00}. After deriving the 
cluster present-day luminosity function, using 1 mag bins ($K_S$ band), we converted the $K_S$ magnitudes to solar masses using values from 
\citet{martins05} for O-type stars. We used \citet{cox00} for stars with type later than B0\,V. The P-DMF is fitted by a  
\citet{kroupa01} IMF and a best fit function. We estimated the cluster total mass by integrating the fitted IMF between 0.10 $M_{\odot}$ 
(log $(M) = -1.00$ dex) and the P-DMF upper mass limit. Our analysis only included the errors associated with the 
fitting to the process. 

\section{Results}\label{resultados}

 \begin{table*}
\caption{Individual extinction and distance estimates for spectroscopically observed stars. Equatorial coordinates, near-infrared magnitudes 
($J$, $H$, and $K_S$), and spectral classification are included.}
\begin{center}
\begin{tabular}{ccccccccc}
\toprule
 ID & RA (J2000) & Dec (J2000)  & $J$ & $H$ & $K_S$ & Spectral type & $A_{K}$ & Distance\\
      & [deg]             & [deg]               & [mag] & [mag] & [mag]  &    &       [mag]        & [kpc]  \\
 \midrule
\multicolumn{9}{l}{VVV CL027} \\
\midrule
s1\_1 & 203.10332 & -62.72932 & 12.246$\pm$0.016 & 11.603$\pm$0.023 & 11.282$\pm$0.002 & O9\,V     & 0.55$^{+0.01}_{-0.01}$ & 6.33$^{+1.85}_{-1.41}$ \\
s1\_2 & 203.10117 & -62.72808 & 12.831$\pm$0.021 & 12.116$\pm$0.037 & 11.795$\pm$0.003 & B1\,V      & 0.59$^{+0.01}_{-0.01}$ & 6.13$^{+1.77}_{-1.39}$ \\
s1\_3 & 203.09920 & -62.72679 & 11.472$\pm$0.016 & 10.782$\pm$0.029 & 10.523$\pm$0.003 & O7--8\,V & 0.56$^{+0.00}_{-0.10}$ & 5.76$^{+1.69}_{-1.28}$ \\
s1\_4 & 203.09803 & -62.72608 & 13.027$\pm$0.016 & 12.373$\pm$0.030 & 12.070$\pm$0.003 & B0--1\,V & 0.55$^{+0.05}_{-0.04}$ & 7.54$^{+2.12}_{-1.71}$ \\
s2\_1 & 203.09905 & -62.72788 & 11.264$\pm$0.037 & 10.448$\pm$0.070 & 10.345$\pm$0.053 & B1--2\,V & 0.53$^{+0.03}_{-0.02}$ & 3.22$^{+0.68}_{-0.39}$ \\
s2\_2 & 203.09809 & -62.72722 & 12.624$\pm$0.014 & 11.920$\pm$0.024 & 11.670$\pm$0.003 & B3\,V      & 0.55$^{+0.10}_{-0.06}$ & 4.55$^{+1.34}_{-1.01}$ \\
\midrule
\multicolumn{9}{l}{VVV CL028} \\
\midrule
01   & 205.09793   &  -61.73414  & 12.141$\pm$0.035 & 11.585$\pm$0.058 & 11.503$\pm$0.003 & B2\,V & 0.40$\pm$0.02 & 5.14$^{+1.46}_{-1.17}$ \\
02   & 205.09552   &  -61.73313  & 13.054$\pm$0.008 & 12.607$\pm$0.016 & 12.370$\pm$0.003 & B3\,V & 0.42$\pm$0.02 & 6.66$^{+1.96}_{-1.48}$ \\ 
\midrule
\multicolumn{9}{l}{VVV CL062} \\
\midrule
01 & 243.028452 & -51.97175 & 14.408$\pm$0.032 & 12.597$\pm$0.050 & 11.544$\pm$0.054 & K0--1\,III  & 1.05$^{+0.03}_{-0.05}$ & 2.62$^{+0.90}_{-0.31}$ \\
02 & 243.033683 & -51.97189 & 14.496$\pm$0.057 & 12.803$\pm$0.063 & 11.872$\pm$0.052 & late G\,III  & $\cdots$ & $\cdots$ \\
03 & 243.038669 & -51.97181 & 10.142$\pm$0.024 &    9.239$\pm$0.024 & 8.847$\pm$0.025 & B5\,V & 0.70$\pm$0.01 & 0.90$^{+0.25}_{-0.21}$ \\
\midrule
\multicolumn{9}{l}{VVV CL088} \\
\midrule
01 & 253.14172   & -44.60115     & 14.413$\pm$0.055 & 12.251$\pm$0.500 & 11.104$\pm$0.500 & O9--B0\,V      & 1.67$\pm$0.01 & 3.28$^{+0.95}_{-0.69}$ \\
02 & 253.14058   & -44.60144     & 14.175$\pm$0.089 & 11.790$\pm$0.500 & 10.502$\pm$0.500 & O9\,V & 1.84$\pm$0.01 & 2.44$^{+0.71}_{-0.55}$ \\
\midrule
\multicolumn{9}{l}{VVV CL089} \\
\midrule
1 & 253.44736     & -43.26425          & 12.640$\pm$0.024 & 10.327$\pm$0.024 & 9.305$\pm$0.021 & O9\,V          & 1.68$\pm$0.02  &  1.51$^{+0.44}_{-0.34}$ \\
2 & 253.44499     & -43.26738          & 13.857$\pm$0.067 & 10.706$\pm$0.051 & 9.344$\pm$0.037 & K2\,III          & 1.79$^{+0.06}_{-0.05}$  & 0.89$^{+0.60}_{-0.23}$ \\
3 & 253.44365     & -43.26912          & 14.336$\pm$0.073 & 11.652$\pm$0.500 & 10.451$\pm$0.500 & K5\,III        & 1.39$^{+0.07}_{-0.01}$  & 3.73$^{+0.23}_{-0.15}$ \\
4 & 253.44280     & -43.27004          & 15.180$\pm$0.070 & 12.949$\pm$0.086 & 11.763$\pm$0.075 & K5\,III        & 1.17$^{+0.06}_{-0.01}$  & 6.14$^{+0.23}_{-0.15}$ \\
\bottomrule
\end{tabular}
\end{center}
\label{data_stars}
\end{table*}


\subsection{Individual clusters}
\begin{itemize}
\item{VVV\,CL027: For this cluster we have the largest set of spectroscopically observed stars. Our spectral classification indicates 
the presence of 2 O-type and 4 B-type dwarfs. We observe them using two slits: slit number 1 contains the stars s1\_1,
s1\_2, s1\_3, and s1\_4; slit number 2 contains stars s2\_1 and s2\_2.

 The spectrum of star s1\_1 presents the \ion{He}{I} 2.06 and 2.11 $\mu m$ lines, and its \ion{H}{I} (4-7) line is clear and broad.
Spectral lines are similar to the O9\,V spectral lines (e.g. O9\,V \object{HD 193322}). Spectra
 s1\_2 clearly shows the \ion{He}{I} 2.11 $\mu m$ and \ion{H}{I} (4-7) lines, similar to B0.5\,V-\object{HD 36960} and 
 B1.5\,V-\object{HD 36959}. We adopted the spectral type B1\,V for this star.
 
 For star s1\_3, we observed \ion{He}{I} at 2.06 $\mu m$, 2.11 $\mu m$, \ion{He}{II} at 2.19 $\mu m$, and the \ion{H}{I} (4-7) lines. For 
 this spectrum, we also detected unidentified features with similar depths. The \ion{N}{III} complex at 2.12 $\mu m$, and the previously 
 mentioned lines fit the O7\,V \object{HD 54662} and the O8\,V \object{HD 48279} spectra. We assumed a O7--8\,V spectral type for this object.
                                
Stars s1\_4, s2\_1, and s2\_2 spectra show the \ion{H}{I} (4-7) and  \ion{He}{I} at 2.11 $\mu m$ lines. For these stars we adopted a
B0--1\,V (star s1\_4), B1--2\,V (star s2\_1), and B3\,V (star s2\_2) spectral type.

 The cluster distance estimate is 6.08$^{+1.56}_{-2.04}$ kpc, and the mass estimate for the cluster is $10^{3.62\pm0.30} M_{\odot}$ (integrating the Kroupa
 IMF fitted until 25 $M_{\odot}$). The \citet{lejeune01} main sequence isochrone fitting to the decontaminated CMD only indicates that the cluster is  younger than 10~Myr. 
 The presence of a O7--8\,V star also sets an upper limit of $\sim$7.0 Myr \citep{meynet00} for the cluster age, in agreement with the pre-main 
 sequence turn-on point observed in the CMD. We estimate a cluster age between 0.7 and 7.0 Myr.

\begin{figure}
\centering
\includegraphics[width=9.5cm]{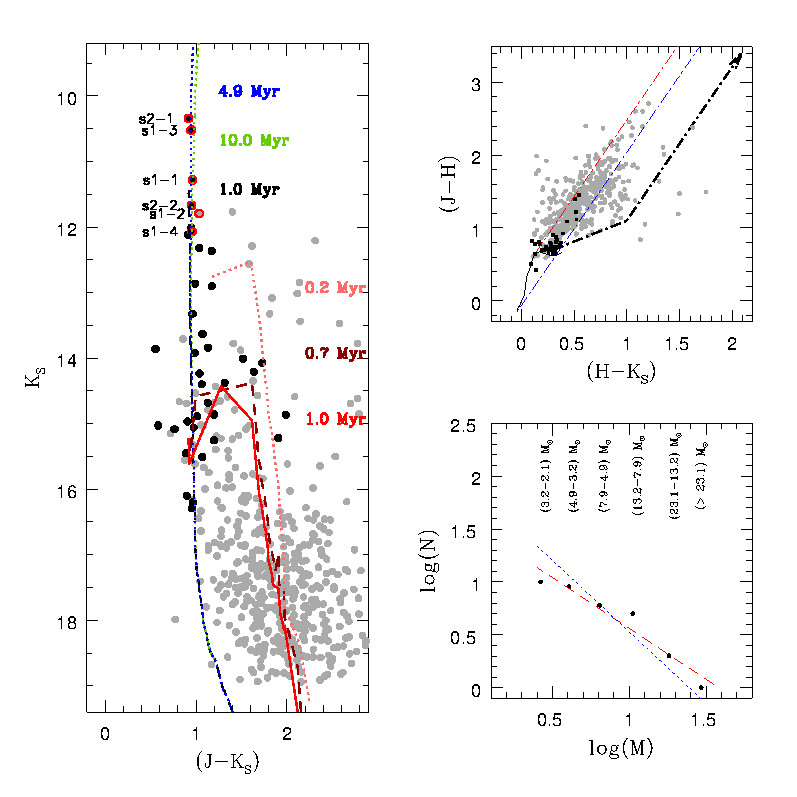}
       \caption{\textit{Left:} VVV\,CL027 field-star decontaminated CMD. The most probable cluster members and field stars are labelled with black 
       and grey symbols, respectively. Numbers indicate the position for the spectroscopically observed stars. Main (1.0, 4.9, and 10.0 Myr, 
       \citealt{lejeune01}) and pre-main sequence (0.2, 0.7, and 1.0 Myr, \citealt{siess00}) isochrones are also shown in the 
       diagram. \textit{Right, top:} VVV\,CL027 field-decontaminated colour-colour diagram (CCD). We use the same symbols as in the CMD to show the most probable 
       cluster members and field stars. The projected position following the \citet{stead09} extinction law of an O8\,V (blue) and a K5\,V (red) star
       are presented in the diagram. The thicker black segmented line shows the expected position for T Tauri stars along the extinction vector.
       \textit{Right, bottom:} VVV\,CL027 P-DMF. The points show the central position in the mass ranges 
            indicated above them, and the segmented red line corresponds to the best fit to the data.}
       \label{diagrams027}
\end{figure}
 }

\item{VVV\.CL028: The two spectroscopically observed stars show early B-dwarf spectral features. The \ion{He}{I} 2.06 $\mu m$ line is not detected
 in any of the spectra, and the \ion{He}{I} 2.11 $\mu m$ line is observed in the spectra of star number 01. The shallower \ion{H}{I} (4-7) line and the 
\ion{He}{I} 2.11 $\mu m$ line indicate that star 01 is earlier than star 02. Both spectra present a clear \ion{H}{I} (4-7) line, but it is not very deep 
(discarding late-B or A-type). We adopted the spectral types B2 and B3\,V for stars 01 and 02, respectively.

The individual distance estimates for the two B\,V observed stars agree with a single distance of 5.90$^{+2.39}_{-1.98}$ kpc.
This cluster is located in the centre of a mid-infrared bubble (Figure \ref{vvvCL028_mir}), which was previously identified and studied by \citet{martins10}. In 
this work, they observed and spectrally classified 18 OB type stars (between O4 and B2.5\,V/III) around the region \object{RCW 79}, 
most of them located around 80$\arcsec$ from the cluster centre. The distance of 4.2 kpc reported by \citet{martins10} allows us to discard
 a possible association between VVV\,CL028 and \object{RCW 79}. Visual inspection of WISE images also indicates that the cluster would not 
 present associated mid-infrared emission, and the individual extinction estimates indicate that the cluster is neither immersed in a cloud nor 
 associated to \object{RCW 79}. 
 
\begin{figure}
\centering
\includegraphics[width=8.5cm]{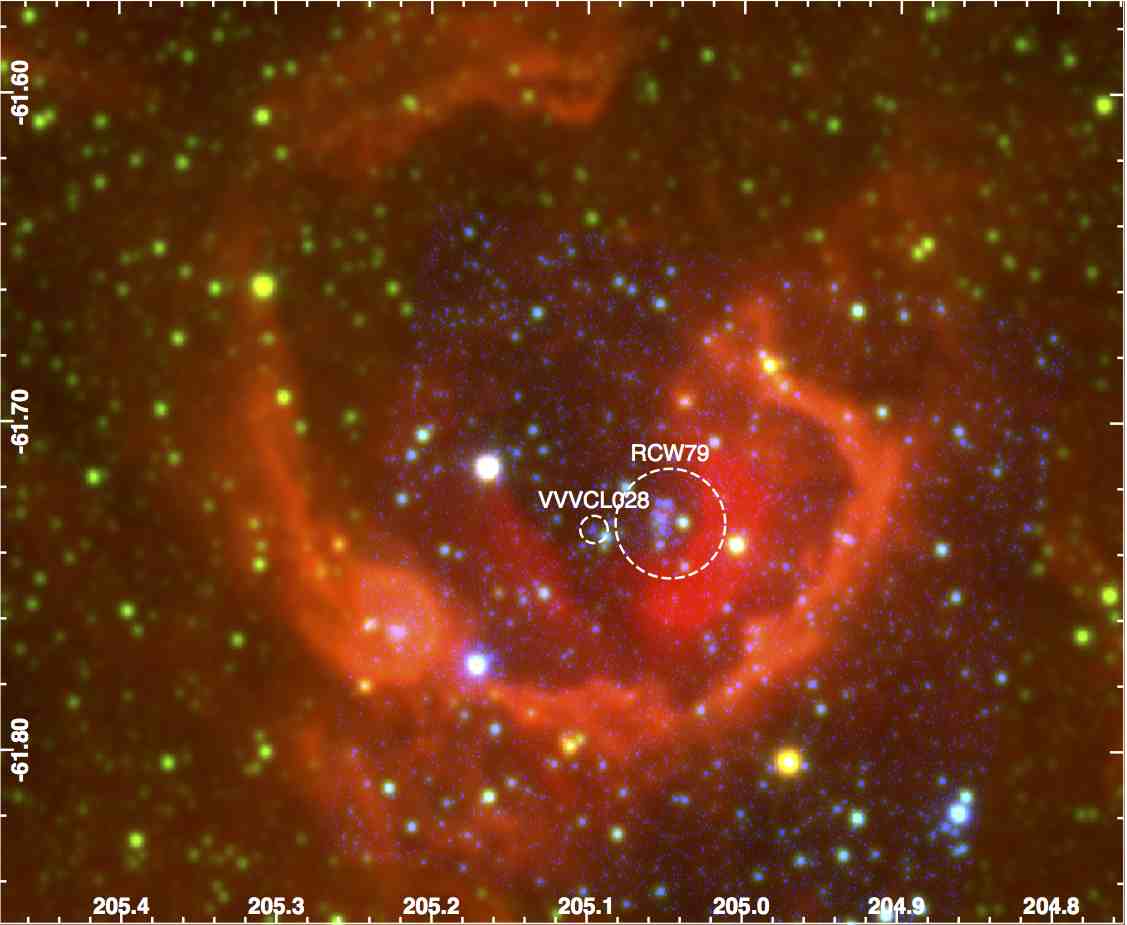}
       \caption{False colour ($K_S$=blue, WISE {\it W}2=green, WISE {\it W}3=red) image for VVV\,CL028. The central yellow circle indicates the position and
        extension of the cluster candidate. North is up, east is left.}
       \label{vvvCL028_mir}
\end{figure}

 By integrating the best fit function to the cluster P-DMF between 0.1 and 11 $M_{\odot}$, we estimated a cluster total mass of
$10^{3.05\pm0.39}$ $M_{\odot}$. We could not identify the cluster main sequence clearly in the CMD (Fig. \ref{diagrams028}), therefore 
an age estimate using isochrone fitting would not yield a reliable result. Considering the B2\,V star as the most massive cluster member, we estimate 
an upper limit  for the cluster age of 20 Myr \citep{meynet00}.
 
 \begin{figure}
\centering
\includegraphics[width=9.5cm]{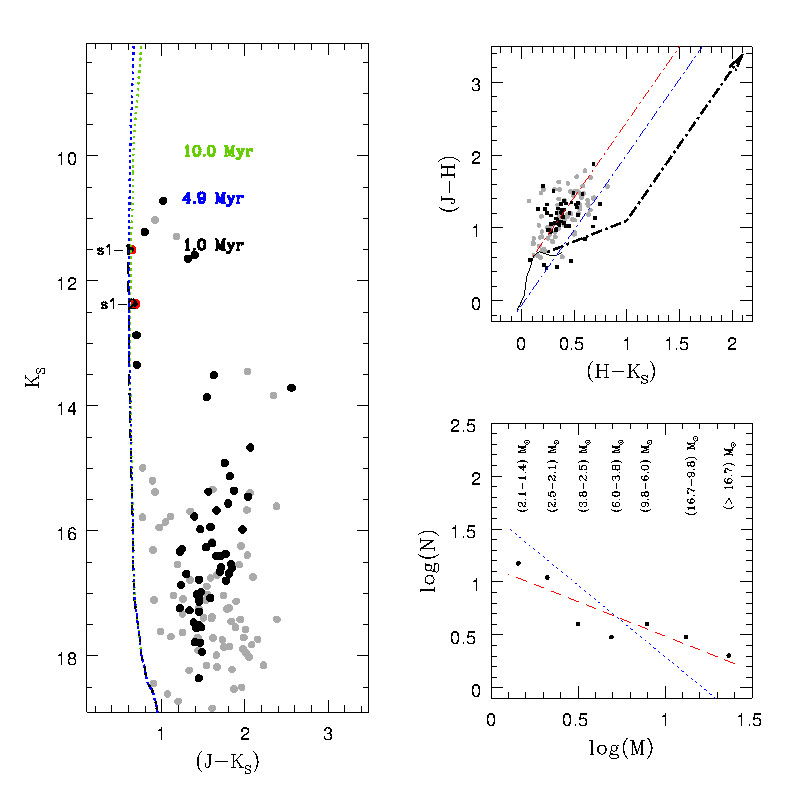}
       \caption{\textit{Left:} VVV\,CL028 field-star decontaminated CMD. Main-sequence (1.0, 4.9, and 10.0 Myr, \citealt{lejeune01}) isochrones are shown 
       in the diagram. For the CMD and CCD we use the same colour code than Figure \ref{diagrams027}. \textit{Right, top:} VVV\,CL028 
       field-decontaminated CCD. \textit{Right, bottom:} VVV\,CL028 P-DMF. The points show the central position in the mass ranges 
       indicated above them, and the lines correspond to the best fit (red, segmented) and the Kroupa IMF fitted (blue, dotted) to the data.}
       \label{diagrams028}
\end{figure}

}
\item{VVV\,CL062: Two of the observed spectra (objects 01 and 02) show the ${}^{12}$CO\,(2,0) band in absorption, indicating a late spectral type. In the
case of spectra 01, it fits K0-1\,III spectra (for example, \object{HR 8694}, \object{HR 8317}, or \object{HR 6299} spectra). The CO band of star number 
02 indicates a late G-giant spectral type. The low signal-to-noise ratio for this spectrum does not allow a more precise classification.

 The spectrum of star 03 clearly presents the \ion{H}{I} (4-7) line in absorption. The depth and shape of this line fits a B5\,V spectral type (between 
 B3\,V \object{HR 5191} and B7\,V \object{HR 3982}). Other features cannot be clearly observed in the spectrum. We adopted this B5\,V distance estimate
 (i.e., 0.90$^{+0.25}_{-0.21}$ kpc) as the cluster distance, but a follow-up (spectroscopic or astrometric) for VVV\,CL062 is necessary for a better characterization 
 of its stellar population).

  The nebulae detected in the WISE W3 image of VVV\,CL062 (Fig. \ref{vvvCL062_mir}) suggest that it is a very young object. The main sequence isochrone 
 fitting in the CMD (Fig. \ref{diagrams062}) indicates than the cluster can be as young as 10 Myr, and the pre-main sequence fitting indicates the presence of 
 objects older than 5.0 Myr. For VVV\,CL062, the P-DMF is reproduced well by the Kroupa IMF. Integrating this function between 0.1 and 15 $M_{\odot}$, 
 we estimate a cluster total mass of $10^{2.03\pm0.14}$ $M_{\odot}$.
 
 \begin{figure}
\centering
\includegraphics[width=7.0cm]{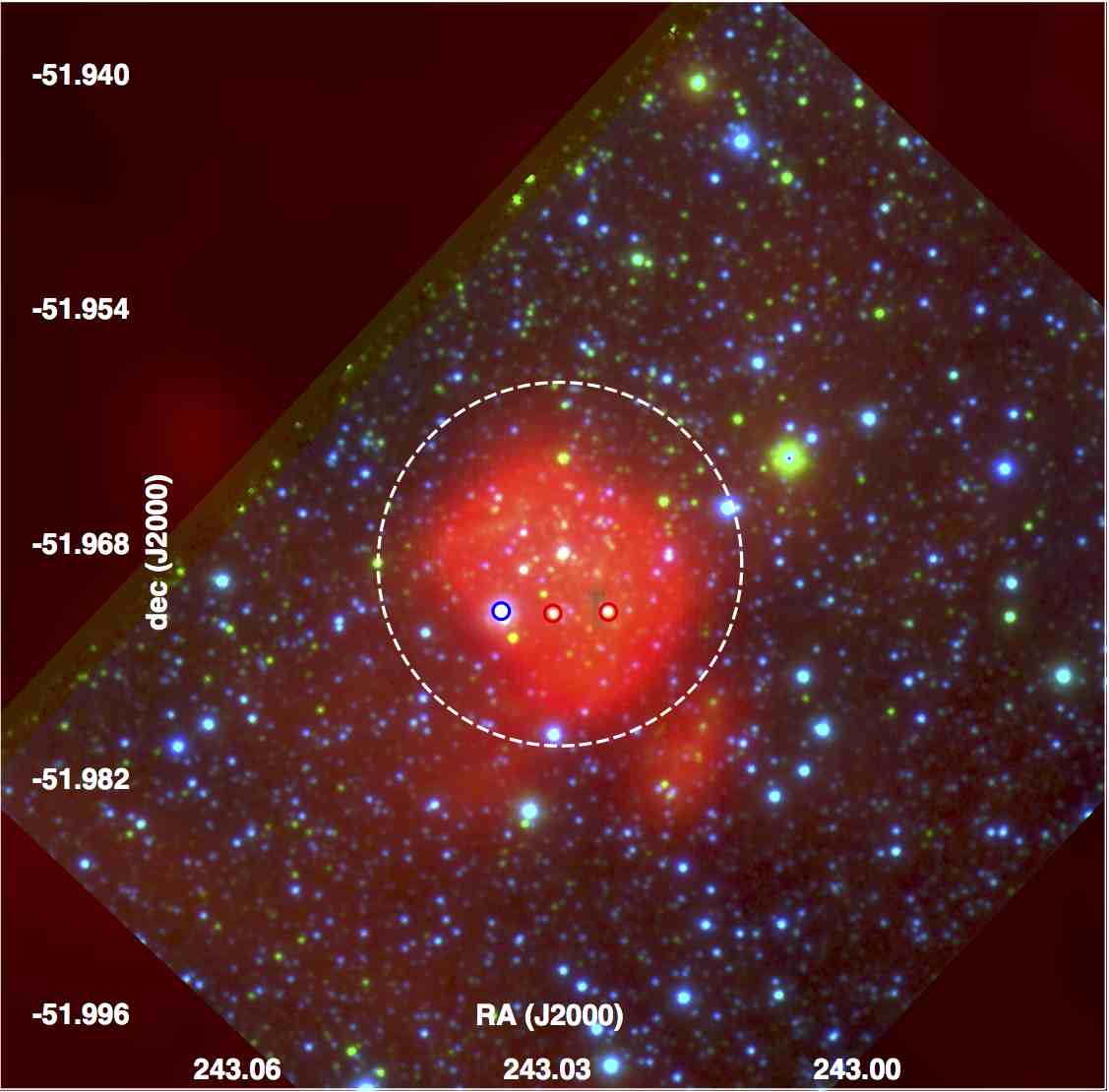}
       \caption{False colour ($J$=blue, $K_S$=green, WISE {\it W}3=red) image for VVV\,CL062. A central white circle indicates the position and
        extension of the cluster candidate. Blue and red small circles show the spectroscopically observed stars. North is up, east is left}
       \label{vvvCL062_mir}
\end{figure}

 \begin{figure}
\centering
\includegraphics[width=9.5cm]{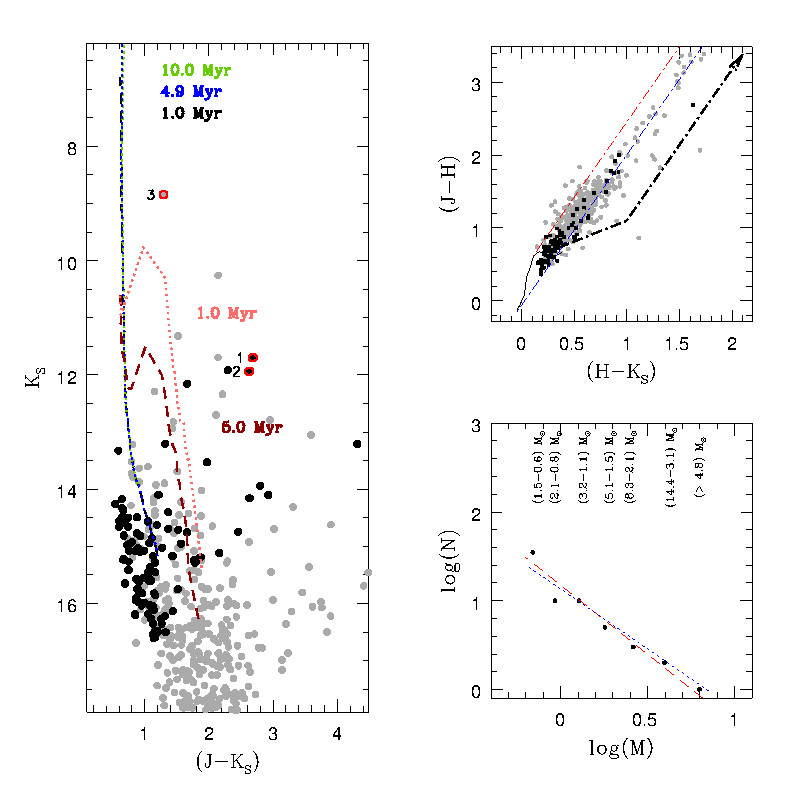}
       \caption{\textit{Left:} VVV\,CL062 field-star decontaminated CMD. Main (1.0, 4.9, and 10.0 Myr, \citealt{lejeune01}) and pre-main sequence (1.0, 
       and 5.0 Myr, \citealt{siess00}) isochrones are shown in the diagram. For the CMD and CCD, we use the same colour code as in Figure \ref{diagrams027} 
       \textit{Right, top:} VVV\,CL062 field-decontaminated CCD. \textit{Right, bottom:} VVV\,CL062 P-DMF. The points show the central position in the mass ranges 
       indicated above them, and the lines correspond to the best fit (red, segmented) and the Kroupa IMF fitted (blue, dotted) to the data.}
       \label{diagrams062}
\end{figure}

}
\item{VVV\,CL088: For this cluster we spectroscopically observed two stars, both very red and bright objects in the CMD (Fig. \ref{diagrams088}). 
Star number 01 spectrum presents both \ion{H}{I} (4-7) and \ion{H}{I} (4-8) in emission. The \ion{He}{I} at 2.06 $\mu m$ line is in emission and \ion{He}{I} at 
2.06 $\mu m$ line is observed in absorption. The absence of the \ion{He}{II} and the depth and shape of the \ion{He}{I} absorption line fit with an 
O9-B0\,V spectral type. The spectrum number 02 is very noise-dominated, and its \ion{H}{I} (4-7) and \ion{He}{II} lines (both below the noise level) fit 
an O9\,V spectrum (e.g. \object{HD 193322}).

\begin{figure}
\centering
\includegraphics[width=8.0cm,angle=0]{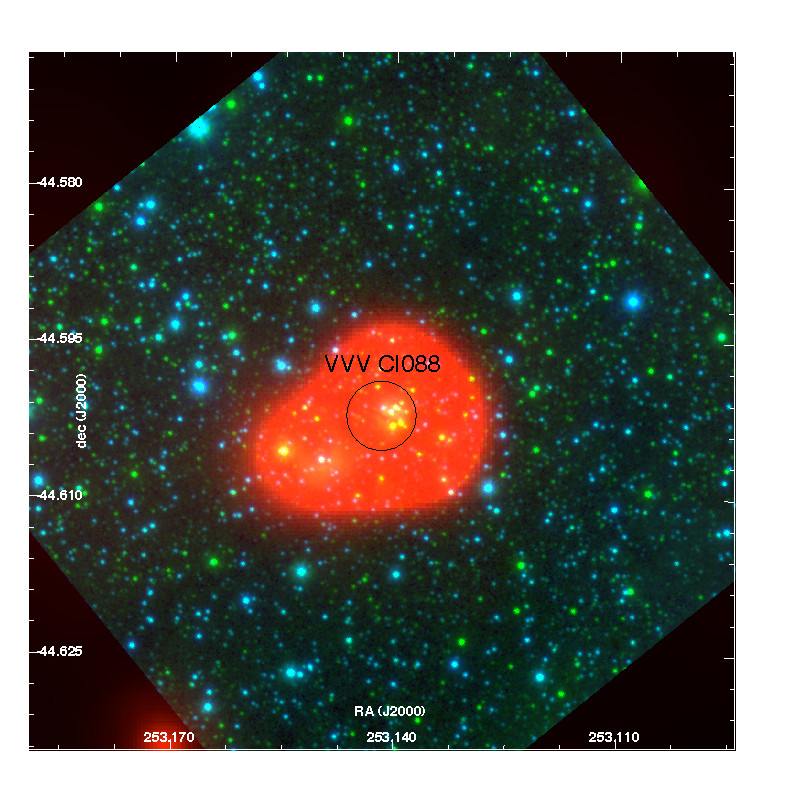}
       \caption{False colour ($J$=blue, $K_S$=green, WISE {\it W}3=red) image for VVV\,CL088. Central black circle marks the first position and
        extension estimated  for the cluster by \citet{borissova11}. North is up, east is left}
       \label{vvvCL088_mir}
\end{figure}

Using the individual distance estimates, we obtain a cluster distance estimate of 2.86$^{+1.16}_{-0.93}$ kpc. The apparent extension of the cluster, based on
the $K_S$ image, is underestimated. Using the WISE mid-infrared images ({\it W}3 band), we find that the cluster extends to at least twice
the size deduced from the $K_S$ image (see Figure \ref{vvvCL088_mir}). Considering a new radius of 0.7\arcmin (the previous estimate by 
\citealt{borissova11} was 12\arcsec), we built the CMD shown in Figure \ref{diagrams088}. The integration of the cluster P-DMF
 obtained from this CMD, between 0.1 and 10 $M_{\odot}$, gives a cluster total mass of $10^{2.47\pm0.26}$ $M_{\odot}$.

 Fitting the isochrone to the CMD gives an estimate of the cluster reddening of $E_{(J-K_S)}=1.8$ mag. We are not able to determine the age via
 isochrone fitting, but the O8\,V cluster population member sets an upper limit of 7.5 Myr \citep{meynet00}. 

 \begin{figure}
\centering
\includegraphics[width=9.5cm]{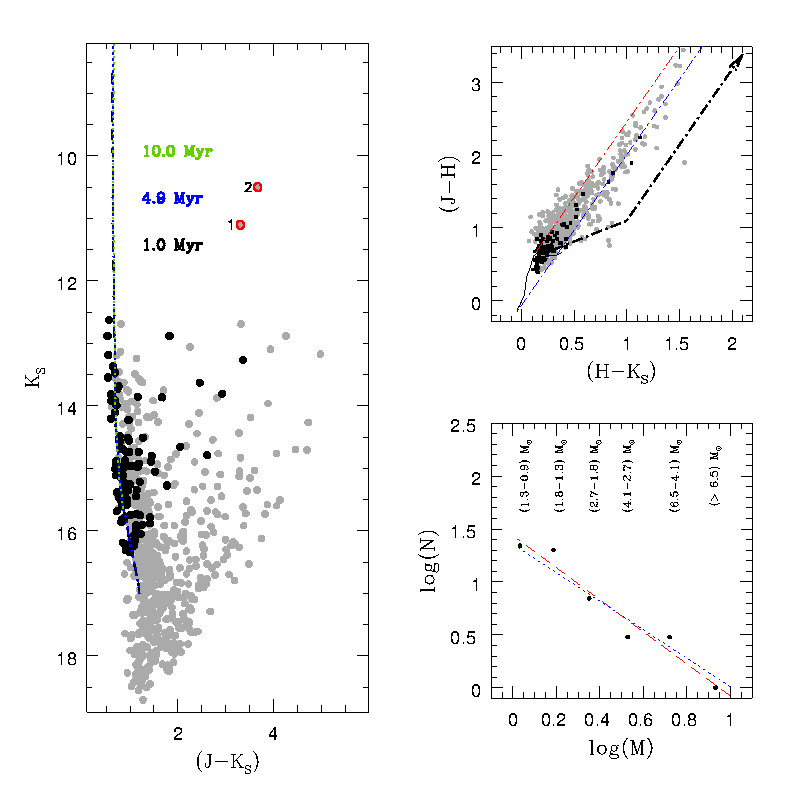}
       \caption{VVV\,CL088 field-star decontaminated CMD with main sequence (1.0, 4.9, and 10.0 Myr, \citealt{lejeune01}) isochrones (\textit{Left}), 
        and CCD (\textit{Right, top}). For the CMD and CCD, we use the same colour code as in
       Figure \ref{diagrams027}. \textit{Right, bottom:} VVV\,CL088 P-DMF. The points show the central position in the mass ranges 
       indicated above them, and the lines correspond to the best fit (red, segmented) and the Kroupa IMF fitted (blue, dotted) to the data.}
       \label{diagrams088}
\end{figure}

}
\item{VVV\,CL089: The spectrum number 01 shows \ion{He}{I} at 2.11 $\mu m$, \ion{H}{I} (4-7), and \ion{He}{II} in absorption. These lines fit the 
O9\,V \object{HD 193322}, and \object{HD 214680} spectra. For this object, we adopted the O9\,V spectral type.
Objects number 02, 03, and 04 present a strong ${}^{12}$CO\,(2,0) band in absorption, which fit the K2-5\,III spectra (for example, \object{HR 6299} 
and \object{HR 1457}). The individual distance estimate of star number 01 was adopted as the cluster distance estimate (1.51$^{+0.44}_{-0.34}$ kpc).

We integrated the Kroupa IMF between 0.1 and 13 $M_{\odot}$, estimating a cluster total mass of $10^{2.78\pm0.13}$ $M_{\odot}$. The main sequence isochrone 
fitting to the CMD gives a cluster reddening of $E_{(J-K_S)}=1.7$ mag. The cluster CMD does not clearly show a pre-main sequence population, making
an age estimation by PMS fitting unreliable. We adopted the upper limit of 9.0 Myr as the cluster age, which was set by the earliest star detected as part of the cluster stellar population 
(i.e., O9\,V).

\begin{figure}
\centering
\includegraphics[width=9.5cm]{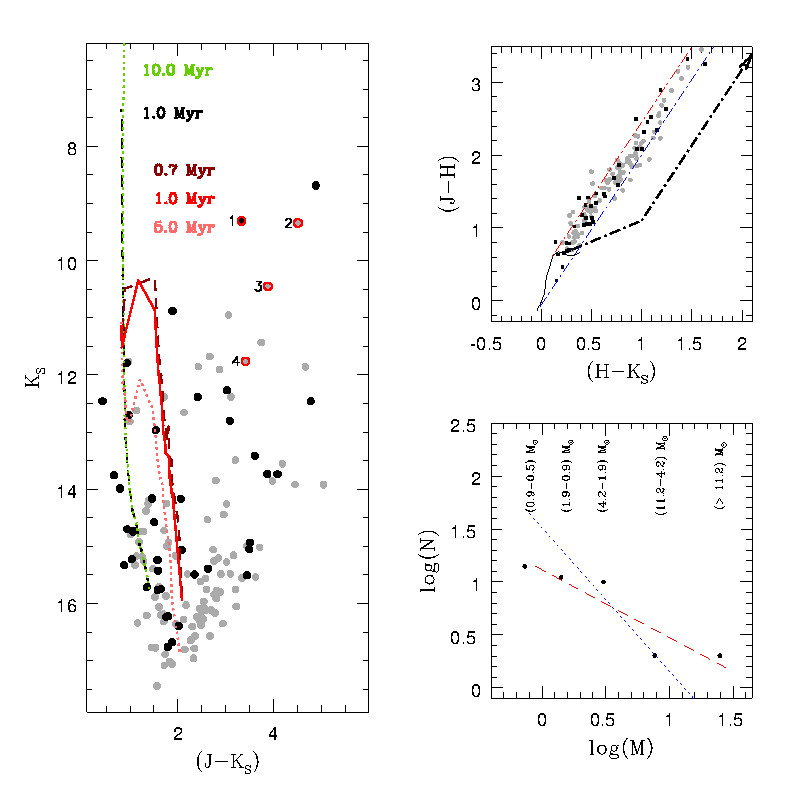}
       \caption{VVV\,CL089 field-star decontaminated CMD (\textit{Left}) and CCD (\textit{Right, top}). For the CMD and CCD we use the same colour code as in
       Figure \ref{diagrams027}. \textit{Right, bottom:} VVV\,CL089 P-DMF. The points show the central position in the mass ranges 
       indicated above them, and the line corresponds to the best fit (red, segmented) to the data.}
       \label{diagrams089}
\end{figure}
}
\end{itemize}

\subsection{Cluster total mass vs. most massive star relation}

\begin{figure}
\centering
\includegraphics[width=9.5cm]{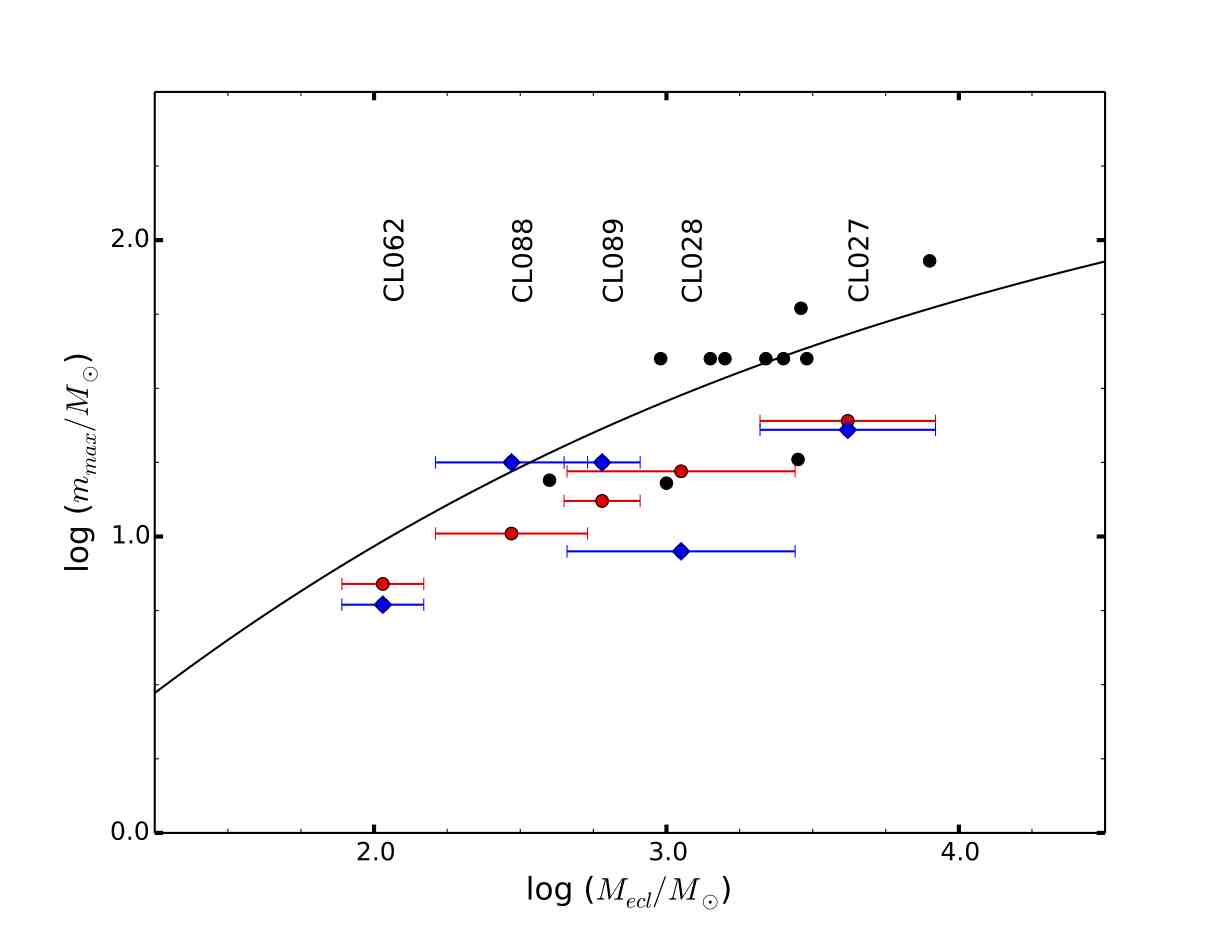}
       \caption{Cluster total mass $M_{ecl}$ versus the mass of the most massive star in the cluster, $m_{max}$. Red circles indicate the position of the stellar mass
       derived from the upper limit of the cluster mass function, and blue rhombi label the positions of the stellar mass estimated from the earliest observed
       spectra. The black solid line shows the ($M_{ecl}$-$m_{max}$) analytical relation \citep{weidner13}. We use black dots to show the mass estimates 
       reported by \citet{chene12,chene13} and \citet{ramirezalegria14}.}
       \label{weidner}
\end{figure}

The relation between the total mass of the cluster $M_{ecl}$ and the mass of its most massive star $m_{max}$ was presented analytically by
\citet{weidner13} (and references therein). This relation implies that the IMF upper mass limit for a determined cluster changes systematically
with the stellar mass of the cluster. This relation was observationally sampled by \citet{weidner10}, using data from very young (less than 5 Myr)
open clusters from the literature. Using our homogeneous sample of physically characterized clusters, we checked the ($M_{ecl}$-$m_{max}$) 
relation, shown in Figure \ref{weidner}. 

From the clusters CMD we understand that our spectroscopically observed stars represent the most massive star adequately in each cluster. There
are no brighter stars with similar colours in the cluster's CMD. In those cases where we can find a brighter cluster probable member, its photometric
features are similar to the spectroscopically observed stars. For a first sample, we selected this set of stars as the $m_{max}$ sample (blue rhombi 
in Fig. \ref{weidner}). The largest discrepancy for this sample is seen for VVV\,CL028, the oldest cluster in our sample (20 Myr). For this cluster, the
mass loss of the earliest stellar members caused by stellar or dynamic evolution can explain the discrepancy. The second sample of $m_{max}$ was 
derived from the upper limit of the cluster P-DMF, determined from the best fit function (red circles, Fig. \ref{weidner}). We 
observe a clear trend in this second sample of parameters, which is similar to the one described by the \citet{weidner13} relation.

 Our data always present lower $m_{max}$ or higher $M_{ecl}$ than expected by the analytical relation. The first scenario can be explained with 
the uncertainty in the spectral type determination or discrepancy in the mass associated to the spectral type. The second scenario is more difficult 
to explain, because our cluster total mass estimate is a lower limit for the cluster mass. Stellar evolution, dynamics, or photometric completeness 
are factors that yield a lower cluster mass than expected. Considering this difference, our sample agrees within the errors the analytical relation
by \citet{weidner13}, supporting an optimal sampling scenario.

\section{Conclusions}\label{conclusiones}
 
 We have presented the physical characterization of five clusters from the \citet{borissova11} catalogue: VVV\,CL027, VVV\,CL028,
 VVV\,CL062, VVV\,CL088, and VVV\,CL089. The clusters contain a spectroscopically observed OB-type population, supported
 by the statistically field-star decontaminated CMD. Using the information from the CMD plus the cluster distance estimates 
 derived from the individual spectroscopic parallax, we derived the cluster P-DMF and the cluster total mass.
 These values range from $(1.07^{+0.40}_{-0.30})\cdot10^2$ $M_{\odot}$ (VVV\,CL062) to $(4.17^{+4.15}_{-2.08})\cdot10^3$ 
 $M_{\odot}$ (VVV\,CL027).
 
  The clusters are very young (1--20 Myr), and most of them present a near-infrared nebulae clearly detected in the VVV-$K_S$ images.
Visual inspection of WISE mid-infrared images reveals a very compact structure and a larger extension than the ones previously estimated 
in two clusters (VVV\,CL062 and VVV\,CL088). The presence of a nebulosity at near and mid-infrared wavelengths supports the 
estimated low age for the clusters.

 Finally, we compared the ($M_{ecl}$ versus $m_{max}$) relation \citep{weidner10,weidner13} with our total cluster mass estimates
 and the most massive cluster member mass value. We found agreement between the analytical relation and our data set. 
The shift between our sample and the ($M_{ecl}$ versus $m_{max}$) analytical relation is expected because of the
 age of our clusters. A comparison with a larger data set, which includes all the mass estimates reported in this paper series, is 
 also presented.
 
 
\begin{acknowledgements}

S.R.A. was supported by the FONDECYT project number 3140605. The VVV Survey is supported by 
ESO, by the BASAL Center for Astrophysics and Associated Technologies PFB-06, by the FONDAP Center for 
Astrophysics 15010003, and by the Ministry for the Economy, Development, and Tourism's Programa Inicativa 
Cient\'{i}fica Milenio through grant IC12009, awarded to The Millennium Institute of Astrophysics (MAS). Support for J.B. 
is provided by Fondecyt Regular No.1120601. P.A. acknowledges the support by ALMA-CONICYT project number 
31110002. M.G. acknowledges support from ESO and Government of Chile Joint Committee 2014. R.K. acknowledges 
partial support from FONDECYT through grant 1130140. J.C.-B. received support from a CONICYT Gemini grant from 
the Programa de Astronom\'{i}a del DRI Folio 32130012.

This publication makes use of data products from the Two Micron All Sky Survey, which is a joint project of the 
University of Massachusetts and the Infrared Processing and Analysis Center/California Institute of Technology, 
funded by the National Aeronautics and Space Administration and the National Science Foundation. The 
Java applet to combine asymmetric errors is available at http://www.slac.stanford.edu/$\sim$barlow/java/statistics1.html.

\end{acknowledgements}

\bibpunct{(}{)}{;}{a}{}{,} 
\bibliographystyle{aa} 
\bibliography{biblio}

\listofobjects

\end{document}